\documentclass[twocolumn,showpacs,preprintnumbers,amsmath,amssymb, superscriptaddress, prb]{revtex4}

\usepackage{graphicx}
\usepackage{dcolumn}
\usepackage{bm}

\usepackage{color}
\definecolor{red}{rgb}{1,0,0}

\definecolor{blue}{rgb}{0,0,1}

\definecolor{green}{rgb}{0,1,0}

\begin{document}
\preprint{APS}
\author {Mariano de Souza}
\email{mariano@rc.unesp.br}
\affiliation{Instituto de Geoci\^encias e Ci\^encias Exatas - IGCE, Unesp - Univ  Estadual Paulista, Departamento de
F\'isica, Cx.\,Postal
178, 13506-900 Rio Claro (SP), Brazil}
\author {Jean-Paul Pouget}
\affiliation{Laboratoire de Physique des Solides, Universit\'e Paris-Sud, CNRS UMR-8502, F-91405 Orsay, France}




\title{Charge-Ordering Transition in (TMTTF)$_2$X Explored via Dilatometry}
%

\vspace{1cm}

\begin{abstract}
Charge-ordering phenomena have been highly topical over the last few
years. A phase transition towards a charge ordered state has been
observed experimentally in several classes of materials. Among them,
many studies have been devoted to the family of quasi-one dimensional organic charge-transfer salts
(TMTTF)$_2$X, where (TMTTF) stands for tetramethyltetrathiafulvalene
and X for a monovalent anion (X = PF$_6$, AsF$_6$ and SbF$_6$). However, the relationship between the electron localization phenomena and the role of the lattice distortion in stabilizing the charge-ordering pattern is poorly documented in the literature.  Here
we present a brief overview of selected literature results with emphasis placed on recent thermal expansion experiments probing the
charge-ordering transition of these salts.

\end{abstract}

\pacs{72.15.Eb, 72.80.-r, 72.80.Le, 74.70.Kn}

\maketitle


\date{\today}

\vspace{3cm}

\section{Introduction}
Charge-ordering (CO) phenomena, also frequently referred to as
charge disproportionation, have been recognized as one of the main
issues in the field of correlated electron systems. In particular,
charge order has gained particular attention in transition-metal
oxides, where, depending on its coupling to spin-ordering, it gives
rise to various exotic phenomena \cite{Imada} such as
ferroelectricity in spiral magnets \cite{Spiral} and magnetoelectric
effects \cite{Magneto}. Also for various series  of molecular
conductors, such as (BEDT-TTF)$_2$X, (TMTTF)$_2$X as well as
(DI-DCNQI)$_2$X, it has been recently recognized that CO plays a key
role in the Physics of these systems \cite{Seo, Kakiuchi}. However,
CO effects observed in these quarter-filled  organic salts is
only a special realization of the so-called 4\,$k_F$ charge density
wave (CDW) instability discovered more than 35 years ago in TTF-TCNQ
\cite{Pouget 1976} (here $k_F$ refers to the Fermi wave vector of the non-interacting 1D electron gas).  Very soon the 4\,$k_F$  CDW was interpreted
within the frame work of the extended Hubbard model as being  due to
a Wigner crystallization caused by the long-range Coulomb repulsions
specific to organic stacking \cite{Hubbard1978, Kondo}. In organic
systems, it is the anisotropy in the hopping transfer integrals,
$t_{ij}$, and mainly the inter-site Coulomb repulsion, $V_{ij}$,
together with the influence from electron-lattice coupling which
give rise to a very rich and diverse Physics;  for a recent review
of all these aspects see e.g. Ref.\,\cite{Pouget2012}. Recently, a new
type of CO with purely ferroelectric character has been reported for
(BEDT-TTF)$_2$X charge-transfer salts \cite{Lunke}. The electronic
properties of molecular conductors have been intensively discussed
in several reviews, see
e.g.\,Refs.\,\cite{Lang4,Lang20a,Ishiguro20,Jerome1991-20,Mckenzie1998-20,
Wosnitza2000-20,Singleton2002-20,Myagawa2004-20,Fukuyama2006-20,Mori2006-20,Kanoda2006-20,Powell2006-20,Dressel2007g7}
and more recently in Ref.\cite{Lebed}.

Here we review literature results on the CO transition occurring in the quasi-one dimensional (1D) (TMTTF)$_2$X Fabre salts. These salts are especially interesting since CO gives rise to
ferroelectricity \cite{Monceau2001-20}. In addition, the concomitant spin-charge decoupling
favors magnetic coupling which in some case leads to a   magnetic ground
state. In this framework it has been proposed \cite{Gianluca} that the interplay
between ferroelectricity and magnetism could lead to multiferroicity. As the Fabre salts are soft materials these electronic phenomena are controlled in a subtle manner by the lattice degrees of freedom. Concerning the various structural
aspects of the so-called Fabre-Bechgaard salts we refer to
Ref.\,\cite{Crystals} and references therein. In this review, emphasis will be placed on high-resolution dilatometric measurements  performed in these salts in the past few years in order to reveal the effects of the  CO transition on the lattice counterpart.

Following this brief introduction which includes the first
section, this brief review is subsequently  divided into eight sections,
as follows:

\begin{itemize}

\item \textbf{Section II}:  an overview of the main theoretical aspects related to the CO transition in quasi-1D organic systems is presented in this section.

\item \textbf{Section III}: this section is dedicated to
the structural aspects of Fabre-Bechgaard
 salts. In particular, we shall emphasize the coupling between the anions and the acceptors which plays a key role in the stabilization of the CO pattern.

\item \textbf{Section IV}: the
generic phase diagram of the Fabre-Bechgaard salts is  first discussed.  This presentation is followed by a
survey of selected literature results related to the CO phase
transition of the Fabre salts.

\item \textbf{Section V}: this section
gives a general presentation of the thermodynamic quantities related to the thermal expansion.

\item \textbf{Sections  VI and VII}: these sections, respectively, present and discuss the high-resolution
thermal expansion results obtained in the (TMTTF)$_2$X's.

\item \textbf{Section VIII}: conclusions, perspectives and an outlook as well as acknowledgements are given here.

\end{itemize}


\section{Theoretical Description of the Charge-Ordering Ground State}\label{COP}
In qualitative terms a CO transition, frequently referred to in the
literature as charge-disproportionation, can be defined as due to  the
self-(re)arrangement of charge carriers into a well-defined
superstructure. This instability is a direct consequence of the
electron-electron repulsions between charges: if the inter-site Coulomb repulsion
$V_{ij}$ is sufficiently strong and of long-range enough, electrons
will reduce the repulsion from each other by taking
equidistant positions.
In 1D, the reduction of the total electrostatic energy of the system is
achieved by the localization of one electron every 1/$\rho$ site, where
$\rho$  is the average number of electron per molecular site; $\rho$
= 1/2 for a quarter-filled system. This localization wave
corresponds to  the 4\,$k_F$ CDW observed in  1D systems (4$k_F$=1/$\rho$ in chain reciprocal wave vector unit).
If the system is metallic above the CO critical temperature ($T_{CO}$), and if $\rho$ is commensurate, the charge localization achieved at CO transition will be then accompanied by a metal-insulator (MI) transition. This situation is, for example, achieved in  (TMTTF)$_2$SbF$_6$ at
$T_{CO,MI}$ $\simeq$ 154\,K \cite{Laversanne84} (see curve 2 in Fig.\,7). In quarter-filled systems, due to the spin-charge decoupling,  the spin-degrees of freedom of the localized electrons remain active below $T_{CO}$. At lower temperatures ($T < T_{CO}$), magnetic driven transitions such as Spin-Peierls (SP)
or antiferromagnetic ordering  can take place.\linebreak

The electronic counterpart of the CO transition has been studied using the extended Hubbard model \cite{Hubbard1978}. The extended Hubbard-Hamiltonian written in second-quantization notation, is
given by:

\vspace*{0.4cm}

\begin{equation}
\emph{H}=\sum_{i,j} t_{ij}a_{i\sigma}^{\dag}a_{j\sigma} +
\sum_i{Un_{i\uparrow}n_{i\downarrow}} +
 \sum_{i,j}V_{ij}n_in_{j}\label{Hubbard}
\end{equation}

\vspace*{0.4cm}

where $t_{ij}$ is the hopping term from site $i$ to site $j$ in the
lattice, $a^\dag_{i\sigma}$ is the creation operator, which creates
an electron on site $i$ with spin $\sigma$ and $a_{i\sigma}$ is the
correspondent annihilation operator; $U$ is the on-site Coulomb
repulsion potential\cite{Onsite}, $n_{i\uparrow}$ and
$n_{i\downarrow}$ refer to the number operator for spin up and down,
respectively, i.e.\ $n_{i\sigma}=a^\dag_{i\sigma} a_{i\sigma}$; and in the last term $n_i$ = $n_{i\uparrow}$ $+$ $n_{i\downarrow}$.

\begin{figure}
\begin{center}
\includegraphics[angle=0,width=0.45\textwidth]{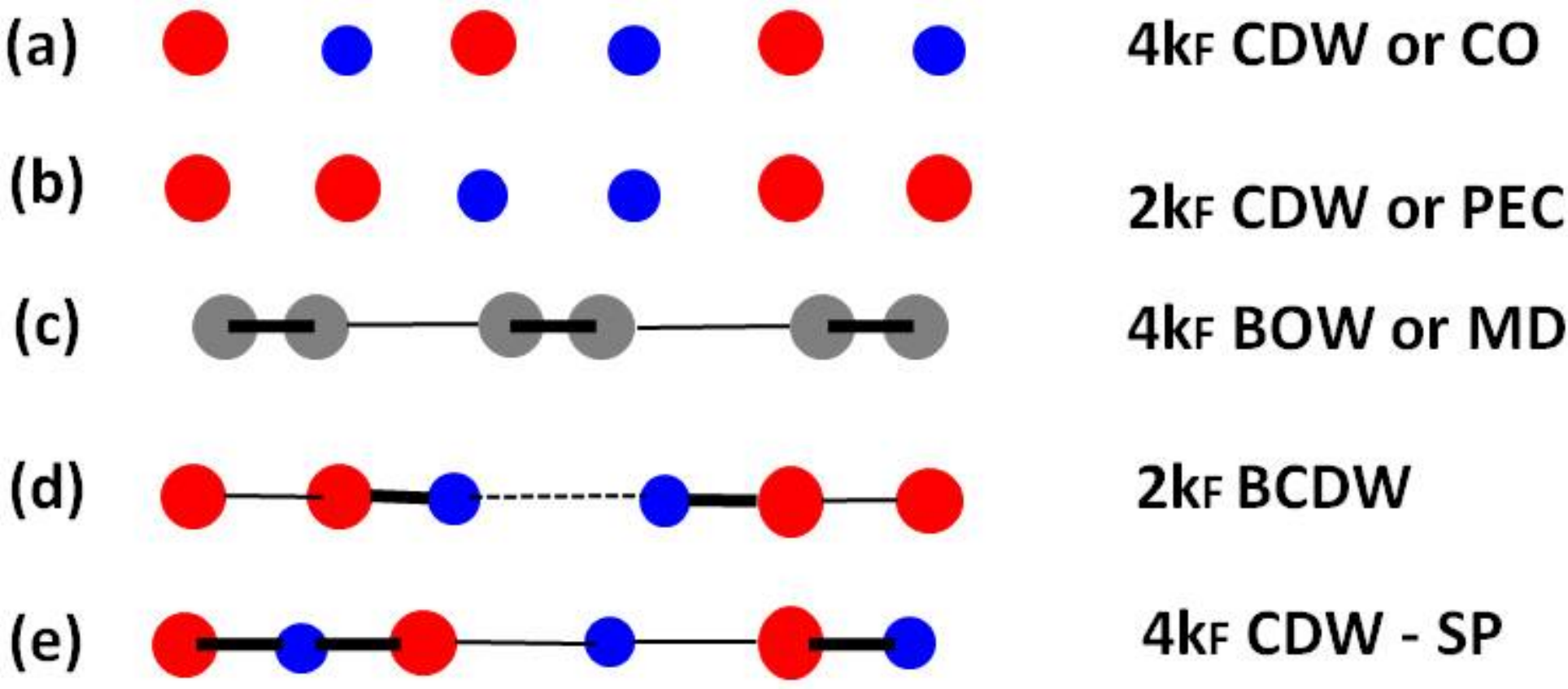}
\caption{\small Schematic representations of charge and bond orders relevant for a quarter-filled band system: (a) 4$k_F$ CDW or CO, (b) 2$k_F$ CDW or PEC, (c) 4$k_F$  BOW or MD \cite{BOW}, (d) 2$k_F$  BCDW and (e) 4$k_F$  CDW-SP. The large red  and small blue dots represent respectively charge-rich
($\rho$ = 1/2 + $\delta$) and charge-poor ($\rho$ = 1/2 $-$ $\delta$) sites. Grey dots represent equally charged ($\rho$ = 1/2) sites. (4$k_F$)$^{-1}$  and (2$k_F$)$^{-1}$  periods correspond to 2 and 4 inter-site distances, respectively. The thickness of the lines linking neighboring sites is proportional to the shortening of the bond distance. In (a) and (b) all the sites are equi-spaced. Note that these ground states present different inversion symmetry except (b) and (d) which have the same symmetry.} \label{COV}
\end{center}
\end{figure}

\begin{figure}
\begin{center}
\includegraphics[angle=0,width=0.47\textwidth]{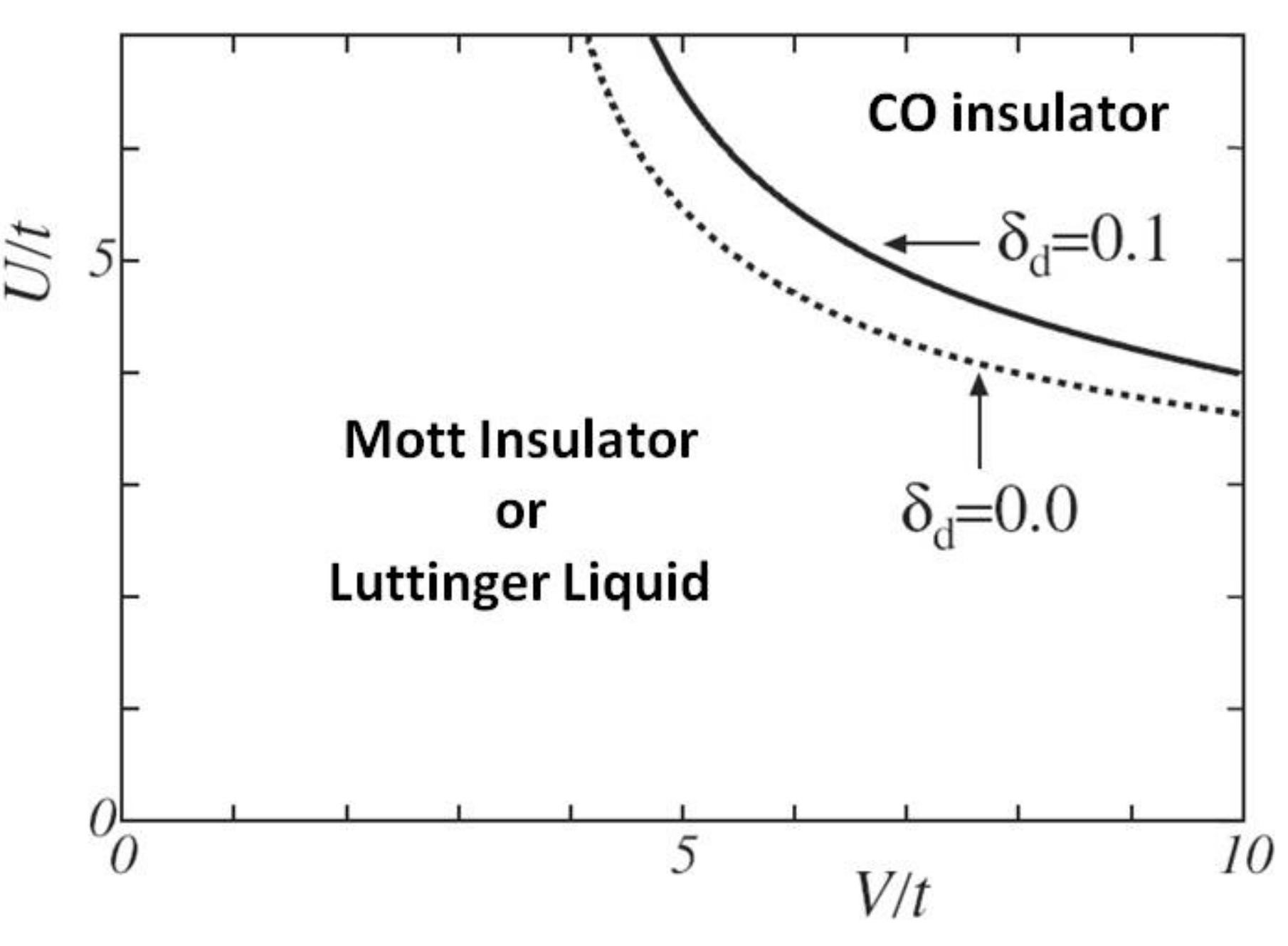}
\caption{\small  ($U/t$, $V/t$) phase diagram of the 1D uniform ($\delta_d$ = 0) and dimerized ($\delta_d$ = 0.1) first neighbor extended Hubbard model at quarter-filling and $T$ = 0\,K. The dashed curve gives in the uniform case the phase boundary between the CO insulator and the Luttinger Liquid.  The continuous line gives in the dimerized case the phase boundary between the CO insulator and the Mott Insulator. Adapted from Ref.\,\cite{Seo}.}\label{PD-CO}
\end{center}
\end{figure}

Already in 1978 Hubbard\cite{Hubbard1978} by using the Hamiltonian (2) without kinetic energy ($t$ = 0) and in absence of double site occupancy ($U$ equal to infinity) was able to discuss Wigner crystals, i.e. 4$k_F$ CDW or CO for $\rho$ between 1 and 1/2. CO requires a convex repulsive potential  $V_{ij}$ between the charges to be stabilized.  In the quarter-filled case, which holds for the (TMTTF)$_2$X family, the 4$k_F$ CDW or CO achieves
a periodicity of twice the
distance between the TMTTF molecules along the chain direction, i.e.\ charge-rich sites alternate with
charge-poor sites in a 1010 pattern, where 1 and 0 refer to
charge-rich and charge-poor sites, respectively (see Fig.\,\ref{COV}  (a)). To be complete, let us remark that a concave repulsive potential $V_{ij}$ between the charges stabilizes the 11001100 charge pattern referred to as a 2$k_F$ CDW or paired electronic crystal (PEC) in the literature (see Fig.\ref{COV}  (b)).
When there are finite on-site Coulomb repulsion $U$ and inter-site hopping $t$,  a very rich phase diagram is obtained because for $t$ larger than the Coulomb repulsions the system is metallic (Luttinger Liquid).  With long-range
Coulomb repulsions restricted to first neighboring sites ($\mid i   -  j \mid$ =  1), exact numerical calculations \cite{Mila, Clay}  show that the quarter-filled system  undergoes a  MI transition achieving the CO ground state shown in Fig.\,\ref{COV}  (a) if $V_{1}$(= $V$ in Fig.\ref{PD-CO}) exceeds  a critical value, $V_c$ (see Fig.\,\ref{PD-CO}). An even richest phase diagram is obtained if second neighbor
interactions ($\mid i   -  j \mid$ = 2) are considered
\cite{Emery88}.  In particular, it is found in this case  a new phase corresponding to the 4$k_F$ bond order wave (4$k_F$ BOW) or Mott Dimer (MD) ground state where the carrier is localized on one bond out of two (Fig.\,\ref{COV}  (c)) \cite{BOW}. When $t$ decreases the 4$k_F$  BOW modulation vanishes at the expense of the 2$k_F$ CDW or PEC.
All these features and more recent calculations including explicitly the electron-phonon coupling\cite{Clay}  show that in reality the CO phase compete with additional ground states exhibiting considerably smaller difference in the charge densities of the charge-rich and charge-poor sites. Some of them superimpose 2$k_F$ BOW or SP distortions at the PEC and CO ground states previously considered, leading to so-called 2$k_F$  BCDW or tetramerized  PEC  (Fig.\,\ref{COV}  (d)) and 4$k_F$  CDW-SP (Fig.\,\ref{COV}  (e)) modulations.

For the (TMTTF)$_2$X salts, where the stack are already dimerized in
the metallic phase,  a modified extended Hubbard model has been
employed (see Ref.\,\cite{Seo} and references therein). In this case,
the Hamiltonian is given by:

\vspace*{0.1cm}

\begin{eqnarray}
\emph{H} &=& t_1\sum_{i\,even,\,
\sigma}(a_{i\,\sigma}^{\dag}a_{i+1\, \sigma} + h.c.) + \nonumber\\
&& t_2\sum_{i\,odd,\, \sigma}(a_{i\,\sigma}^{\dag}a_{i+1\, \sigma} +
h.c.) + \nonumber\\ &&  U\sum_i{n_{i\uparrow}n_{i\downarrow}} +
V\sum_{i}n_in_{i+1}\label{ExtendedHubbard}
\end{eqnarray}


where   $t_1$ = $t (1 - \delta_d)$  and  $t_2$ = $t (1 + \delta_d)$  are, respectively, the inter- and
intra-dimer transfer integrals and  $V$ is the nearest-neighbor
Coulomb repulsion.  The phase diagram of this Hamiltonian is compared in Fig.\,\ref{PD-CO} to the one previously obtained for the uniform quarter-filled repulsive chain.  The dimerization of the chain creates favorable conditions for a 4$k_F$ BOW charge localization (Mott Insulator). As this latter has a different inversion symmetry as the CO, this destabilizes the 4$k_F$  CDW instability and thus enhance the critical $V_c$ necessary to achieve the  CO ground state \cite{Tsuchiizu2001}. In the Fabre salts, because of the incipient stack dimerization a 4$k_F$  BOW progressive charge localization is generally observed at a crossover temperature $T_{\rho}$  larger than the critical temperature $T_{CO}$ achieving the CO (see Section IV).

Figure 1 shows various $T$ = 0\,K intra-chain charge configuration resulting from the minimization of various models of the 1D interacting electron gas. The experimentally relevant 3D ground state must be stabilized by inter-chain coupling. Such inter-chain coupling involves Coulomb coupling between charges sitting on neighboring chains or effective interactions between charges mediated by lattice deformations. The effect of inter-chain Coulomb coupling has been recently considered in the Fabre salts \cite{Gianluca}. The consideration of the lattice degrees of freedom requires the inclusion of electron-phonon coupling terms in the Hamiltonian as done, for example, in Ref.\,\cite{Clay}. The stabilization of the various charge modulation patterns shown in Fiq.\,1 requires a coupling to the intra-molecular lattice modes or to the anions (see Ref.\,\cite{Riera2001}), while the various bond distance modulation patterns require a coupling to the inter-molecular lattice modes (these aspects are briefly reviewed in Ref.\,\cite{Pouget2012}). In the Fabre salts both charge modulation and stack deformation (this latter leading to a deformation of the cavity where the anion is located)  will be coupled to the anion position and orientation and thus transmitted to the lattice parameters via the large phononic Gr\"uneisen parameter of the anion (see section V).

Concluding this Section, it is useful to
mention that evidence for a CO phase has been also reported in many quarter-filled organic compounds other than the Fabre salts. For example, first evidences of a phase transition to a CO state were found
by means of infra-red (IR)
spectroscopy  in the
quasi-2D conductor $\alpha$-(BEDT-TTF)$_2$I$_3$
\cite{Moldenhauer9a1}, and via NMR studies in the quasi-1D conductor
(DI-DCNQI)$_2$Ag \cite{Hiraki9a2}.
However, to our knowledge, the very first report of a CO transition
combining both NMR and structural studies was performed in the (TMP)$_2$X-CH$_2$Cl$_2$ (X = PF$_6$, AsF$_6$) system \cite{Ilakovac}.

\section{Crystal Structure of the Fabre-Bechgaard Salts}\label{Crystal Structure of TMTTF}

\begin{table}[htb]
\centering \small
\begin{tabular}{|c|c|c|c|}
\hline\hline \textbf{X} & \textbf{PF$_6$} & \textbf{AsF$_6$} & \textbf{SbF$_6$} \\
\hline\hline
  $a$(\AA) & 7.146 & 7.178 & 7.180\\ \hline
  $b$(\AA) & 7.583 & 7.610 & 7.654\\ \hline
  $c$(\AA) & 13.218 & 13.317 & 13.51\\ \hline
  $\alpha$ ($^0$) & 82.69 & 82.03 & 81.24\\ \hline
  $\beta$ ($^0$) & 84.87 & 95.75 & 83.42\\ \hline
  $\gamma$ ($^0$) & 72.42 & 107.11 & 74.00\\ \hline
  $V$(\AA$^3$) & 676 & 687 & 702.9\\ \hline
    $d_1$ (\AA) & 3.66 & 3.64 & 3.64\\ \hline
    $d_2$ (\AA) & 3.52 & 3.54 & 3.52\\ \hline\hline
\end{tabular}
\caption{\small Lattice parameters of the (TMTTF)$_2$X salts at
room temperature for X = PF$_6$ \cite{Gallois20g5}, AsF$_6$
\cite{Chasseaug20g6} and SbF$_6$ \cite{Furukawa-2005}, together with the inter- and intra-dimer distances $d_1$ and $d_2$.}
 \label{TMTTFUC}
\end{table}

\begin{figure}
\centering
\includegraphics[angle=0,width=0.48\textwidth]{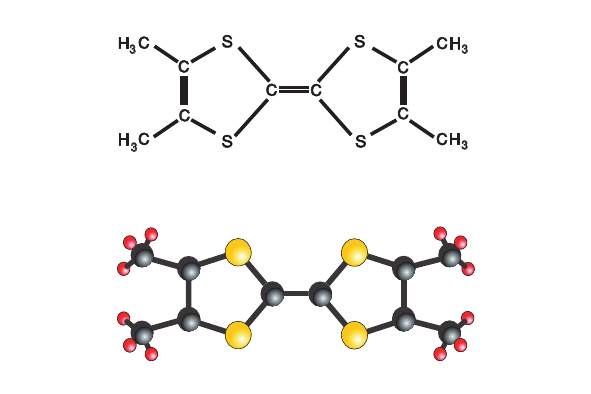}
\caption{\small The (TMT\emph{C}F) molecule, which is the basic
entity building the Fabre-Bechgaard salts.
\emph{C} (yellow circles in the bottom part of the figure) stands for S or Se atoms. The TMTTF molecule is represented in the upper part of the figure. Taken from
\cite{Wosnitza2000-20}.} \label{TMTTF}
\end{figure}

\begin{figure}
\begin{center}
\includegraphics[angle=0,width=0.515\textwidth]{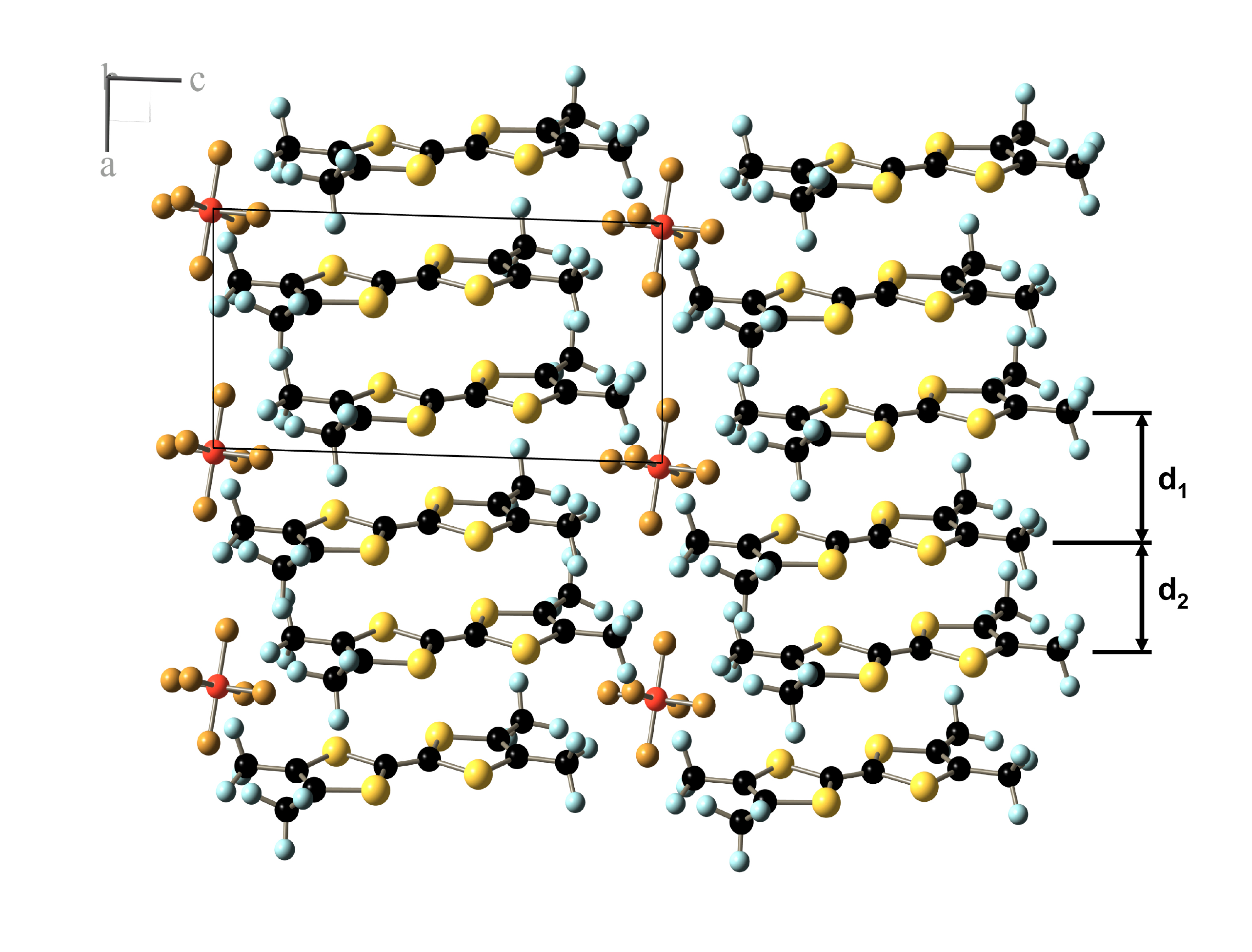}
\caption{\small Projection of the
crystallographic structure of (TMTTF)$_2$X in the \emph{a}-\emph{c} plane. $d_1$ and $d_2$  are the interdimer and the intradimer
spacing, respectively, with $d_1$ $>$ $d_2$. The unit cell is
outlined in black. Taken from Ref.\,\cite{Thesis}.} \label{TMTTFstr}
\includegraphics[angle=0,width=0.515\textwidth]{PF6_bc_finale.pdf}
\end{center}
\caption{\small Structure of
(TMTTF)$_2$X projected in the \emph{b}-\emph{c} plane. The anion (X = PF$_6$, AsF$_6$
or SbF$_6$) is shown in an octahedron representation. Red lines
indicate the shortest distance between the F and S atoms. Note that each anion experiences two short S -- F distances related by inversion symmetry in the RT structure and only one short S -- F distance in the CO state. The unit
cell is outlined by yellow lines.  Taken from Ref.\,\cite{Thesis}.} \label{TMTTFstr_bc}
\end{figure}

The building block of the Fabre-Bechgaard family of organic
conductors is the\linebreak (TMT\emph{C}F) molecule (shown in
Fig.\,\ref{TMTTF}); here \emph{C} stands for the chalcogene S or
Se atoms.  Irrespective of the counter ion, all
(TMT\emph{C}F)$_2$X salts crystallize at room temperature (RT)  in the centrosymetric triclinic structure
P$\bar{1}$ with two donor molecules and one anion in the unit
cell, cf.\,Fig.\,\ref{TMTTFstr}.

As can be seen from Figs.\,\ref{TMTTFstr} and \ref{TMTTFstr_bc}, the
quasi-planar (TMT\emph{C}F) molecules are arranged in a zig-zag
configuration forming stacks along the \emph{a}-axis, where
the highest electrical conductivity is observed in these materials.
The stacks form layers in the \emph{a}-\emph{b} plane, which
themselves are separated by anions along the \emph{c}-axis, so that
shortest separations between  sulfur (for the TMTTF family) or
selenium (for the TMTSF) atoms located on neighboring molecules are
in the \emph{a}-\emph{b}  plane. Figs.\,4 and 5 show more precisely
that terminal methyl groups of successive (\emph{a}, \emph{b}) TMTCF
layers delimit along the inter-layer direction \emph{c$^*$} soft
cavities filled by anions. In such cavities the anion experiences a
quite symmetric environment due to the presence of six closest
methyl groups \cite{Kistenmacher1984}.

The anions can be classified according to their symmetry. For
example, X = Br (spherical); PF$_6$, AsF$_6$ and SbF$_6$
(octahedral) are centrosymmetrical anions, while BF$_4$,ClO$_4$,
ReO$_4$ (tetrahedral) and SCN (linear) are non-centrosymmetrical. As
we shall see below, the counter ions are not only responsible for
making the charge balance of the (TMT\emph{C}F)$_2^+$ donor
molecule, but their structural degrees of freedom  also influences
dramatically the electronic properties of the Fabre-Bechgaard salts.
At RT the anions, even centro-symmetric ones such as
PF$_6$, as well as the methyl groups, are subject to considerable
thermally activated motions. In this regard, NMR measurements in (TMTSF)$_2$PF$_6$
\cite{Scott1982,McBrierty1982,Mc1982} and (TMTTF)$_2$SbF$_6$
\cite{Furukawa-2005,Yu2004-20}  show that both anion and
methyl-group rotational disorders are progressively removed upon
cooling. In (TMTSF)$_2$PF$_6$ the classical motion of the PF$_6$ and of the
methyl groups probed by NMR stops below about 70\,K \cite{Mc1982}
and 55\,K \cite{Scott1982}  respectively, but quantum tunnelling of
the methyl groups remains below these temperatures. The 4\,K
structural refinements of (TMTSF)$_2$PF$_6$
\cite{Gallois86,Foury2013}  and (TMTTF)$_2$PF$_6$
\cite{Gallois20g5,Granier1988} show that each anion locks its
orientation by establishing two short F-Se contact distances and
four F...H-CH$_2$ bonds with four neighboring TMTCF molecules
located nearly in a plane perpendicular to \emph{a}. In
(TMTSF)$_2$PF$_6$,  a cooperative locking between anions and methyl
groups occurs at $T$ $\approx$ 55\,K \cite{Scott1982,Foury2013} via
the formation of H bonds between these two entities. Non
centro-symmetric anions X, located in similar methyl group cavities
are also subject to an orientational disorder at RT.
The orientations of these anions order upon cooling in a staggered way through a
symmetry breaking phase transition which stabilizes a
superstructure where some of the high temperature lattice parameters
are doubled \cite{Pouget96}.

In the TMTSF salts, the interstack interaction is stronger than in
the TMTTF salts. This is because the shortest Se-Se interstack
distances in the TMTSF family are less than twice the van der Waals'
radii of the Se atoms, whereas in the TMTTF family the corresponding
shortest S-S distances are more than twice the van der Waals' radii
of the sulfur atoms. For this reason TMTTF salts are more 1D than
the TMTSF salts. The ratio of transfer integrals  in the \emph{a},
\emph{b}, \emph{c} directions is roughly $t_a$ : $t_b$ : $t_c$ = 100
: 10 : 1, cf.\,\cite{Ishiguro20}.

The unit cell parameters of the salts (TMTTF)$_2$X (X = PF$_6$,
AsF$_6$ and SbF$_6$) investigated here are listed in
Table\,\ref{TMTTFUC}.

An important aspect of the structure of these materials is that the
donors are not equally spaced in the stack direction. There are two
distinct distances between consecutive molecules in the same stack,
as indicated by $d_1$ and $d_2$ in Fig.\,\ref{TMTTFstr} and Table I,
with $d_1$ $>$ $d_2$.
This feature gives rise to the formation of dimers of
TMT\emph{C}F molecules along the stacks. Dimerization is crudely
defined as 2($d_1$ $-$ $d_2$)/\emph{d}, with \emph{d} being the
average distances between TMT$C$F molecules along the stacks, cf.\,Fig.\,\ref{TMTTFstr}. A more elaborated definition of the dimerization
parameter $\delta_d$, used in the Hamiltonian (21), replaces the $d_i$ by the difference of intra- and inter-dimer
transfer integrals \cite{Pouget96}. Stack dimerization, more
pronounced in the TMTTF family, depends also upon the geometry of
the anion X. Dimerization plays a very important role in the
physical properties of  the most 1D (TMTTF)$_2$X system by inducing
additional 4\,$k_F$ Umklapp scattering terms between electrons
\cite{Emery82}. Without taking into account dimerization effects,
the band is strictly quarter-filled in terms of holes, while an effective
half-filled band has to be considered if dimerization is relatively important. Upon cooling, dimerization causes in a 1D electron gas, via the above-quoted
4\,$k_F$ Umklapp scattering term, a crossover to a Mott-Hubbard
charge localized state of one hole per dimer (MD state shown in Fig.\,\ref{COV} (c)). 
This crossover is clearly evidenced by the onset of an activated
conductivity below $T_{\rho}$  $\sim$ 100 -- 250\,K, whose value
significantly depends upon the anion X,  see Fig.\,7
\cite{Laversanne84,Coulon82}. This localized state is
referred to as loc in the phase diagram shown in Fig.\,\ref{Dressel}.

Finally, let us note that with two TMTTF molecules per
period \emph{a},  the formation of a CO pattern  on a preexisting stack dimerization  will not change the
translational symmetry in the stack direction of the Fabre-Bechgaard salts. However, as the MD and CO orders have a different inversion symmetry (see Fig.\,\ref{COV} (c) and (a), respectively) the stabilization of a CO on a preexisting DM order  should lead to a symmetry lowering by removing all the inversion centers in the stack direction.
In these conditions the CO induces a dielectric polarization of the stack.

\section{ Phase diagram of the Fabre-Bechgaard salts}

\textbf{A. The generic phase diagram}

The era of organic superconductors begins with the observation of
superconductivity in pressurized (TMTSF)$_2$PF$_6$ by D. J\'erome
\emph{et al.} \cite{Jerome82} at the end of the 70's. Since then for all the anions (at the exception of X = NO$_3$) both the Bechgaard and Fabre salts were found superconductor under pressure. (TMTSF)$_2$ClO$_4$ is the only ambient pressure superconductor.
The Fabre and Bechgaard salts present a very rich phase diagram under pressure (see Fig.\,6).  The electronic properties of (TMTTF)$_2$X are more 1D than those of (TMTSF)$_2$X. This feature together with the fact that TMTTF salts exhibit stronger electron repulsions than TMTSF salts lead in the low pressure range of the phase diagram of the Fabre salts to a charge localized behavior (below $T_{\rho}$ - see Fig.\,7) generally followed by a CO transition. The strong inter-chain coupling present in the Bechgaard salts induces a 1D to 2D then to 3D deconfinement of the electron gas upon cooling (see Fig.\,6). Superconductivity is achieved in the 3D delocalized state.
Here  the discussion  will be
mainly focused on the properties of (TMTTF)$_2$X family which
exhibits CO transitions. It is well-known that quasi-1D conductors
are inherently more susceptible to sustain  density wave
instabilities than 2- or 3-D conductors. Due to the combination of
the 4$k_F$ charge localization instability and of the 2$k_F$ nesting
instability of the correlated electron gas in 1D, together with the
interplay of substantial electron-phonon interactions, the phase
diagram of the Fabre-Bechgaard salts covers a rich variety of ground
states.

\vspace{0.4cm}

\textbf{1. Centro-symmetric anions}

\begin{figure}
\begin{center}
\includegraphics[angle=0,width=0.48\textwidth]{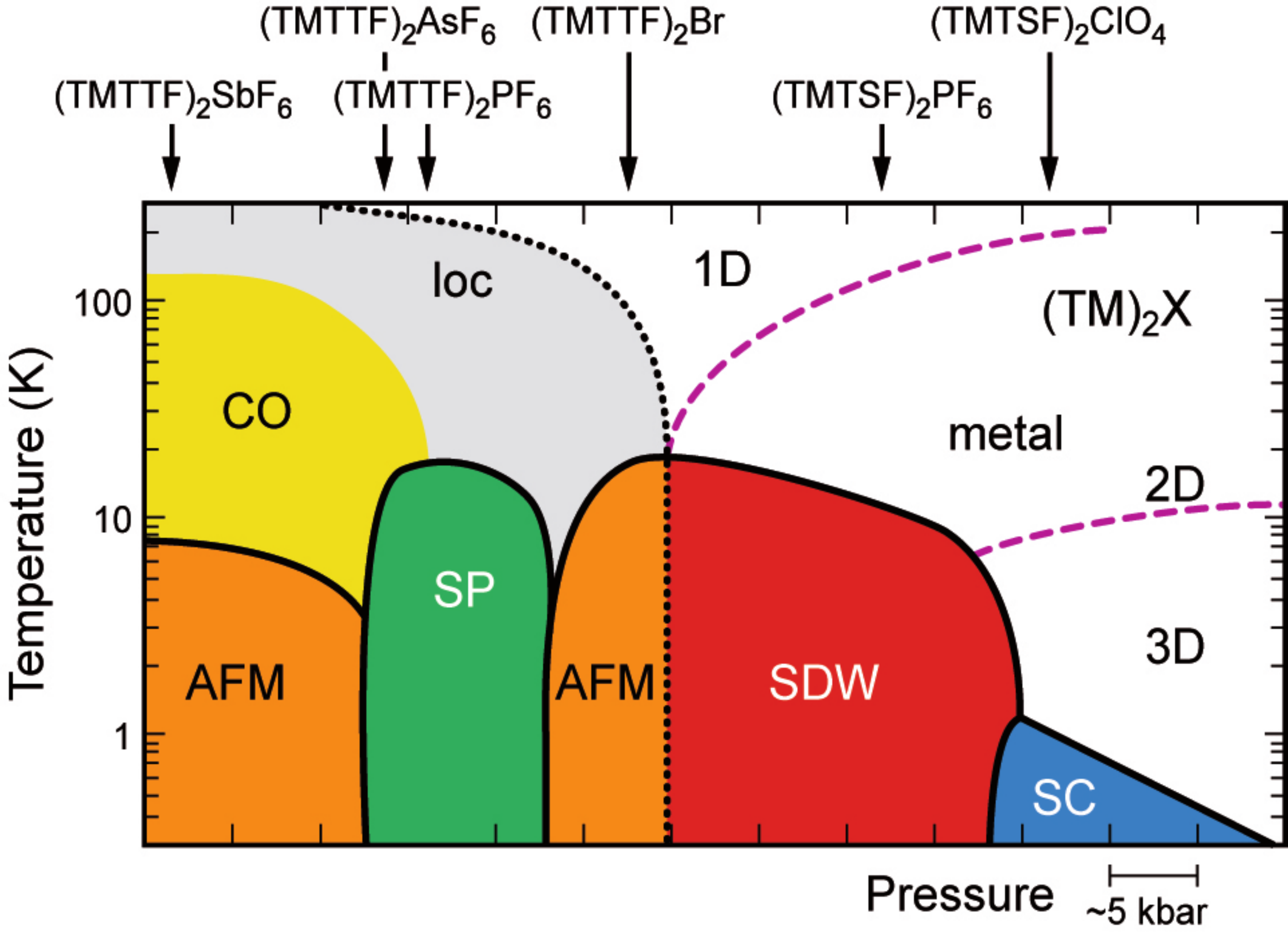}
\end{center}
\caption{\small Generic $P$ - $T$ phase diagram of the
Fabre-Bechgaard salts. The position of the different compounds at
ambient pressure is indicated by arrows. The various ground states
are Charge-Ordering (CO), Spin-Peierls (SP), Antiferromagnetic
(AFM), Spin-Density Wave (SDW), and Superconductivity (SC).  The
crossover to a charge localized state (loc) is indicated by the
black dashed line on the left side of the phase diagram.
Deconfinement crossover when the coherence of the electron gas
passes from 1D- to 2D- or even 3D is indicated by the red dashed
lines on the right side of the diagram.  For details on the pressure
dependence of $T_{CO}$ for the X = SbF$_6$ and AsF$_6$ salts, see
Fig.\,\ref{NMRunderP}.   Taken from Ref.\,\cite{Dressel2007g7}.}
\label{Dressel}
\end{figure}

The phase diagram of the (TMT\emph{C}F)$_2$X salts where X is centro-symmetric has been extensively studied. The most recent version of their generic $P$ -- $T$ phase diagram is shown in Fig.\,\ref{Dressel} (note that in the earlier versions of
the phase diagram \cite{Jerome1991-20,Bourbonnais 97,Bourbonnais 98}
the CO phase was not established). As Fig.\,\ref{Dressel} highlights, the
ground states can be
tuned by anion substitution, exchange of the chalcogen atom of the
donor molecule (\emph{C} = Se or S) and/or external pressure
application. The phase diagram represented in Fig.\,\ref{Dressel} shows in particular that the ground states of the TMTTF family evolve towards those of the TMTSF family by applying pressure.

\begin{figure}
\begin{center}
\includegraphics[angle=0,width=0.47\textwidth]{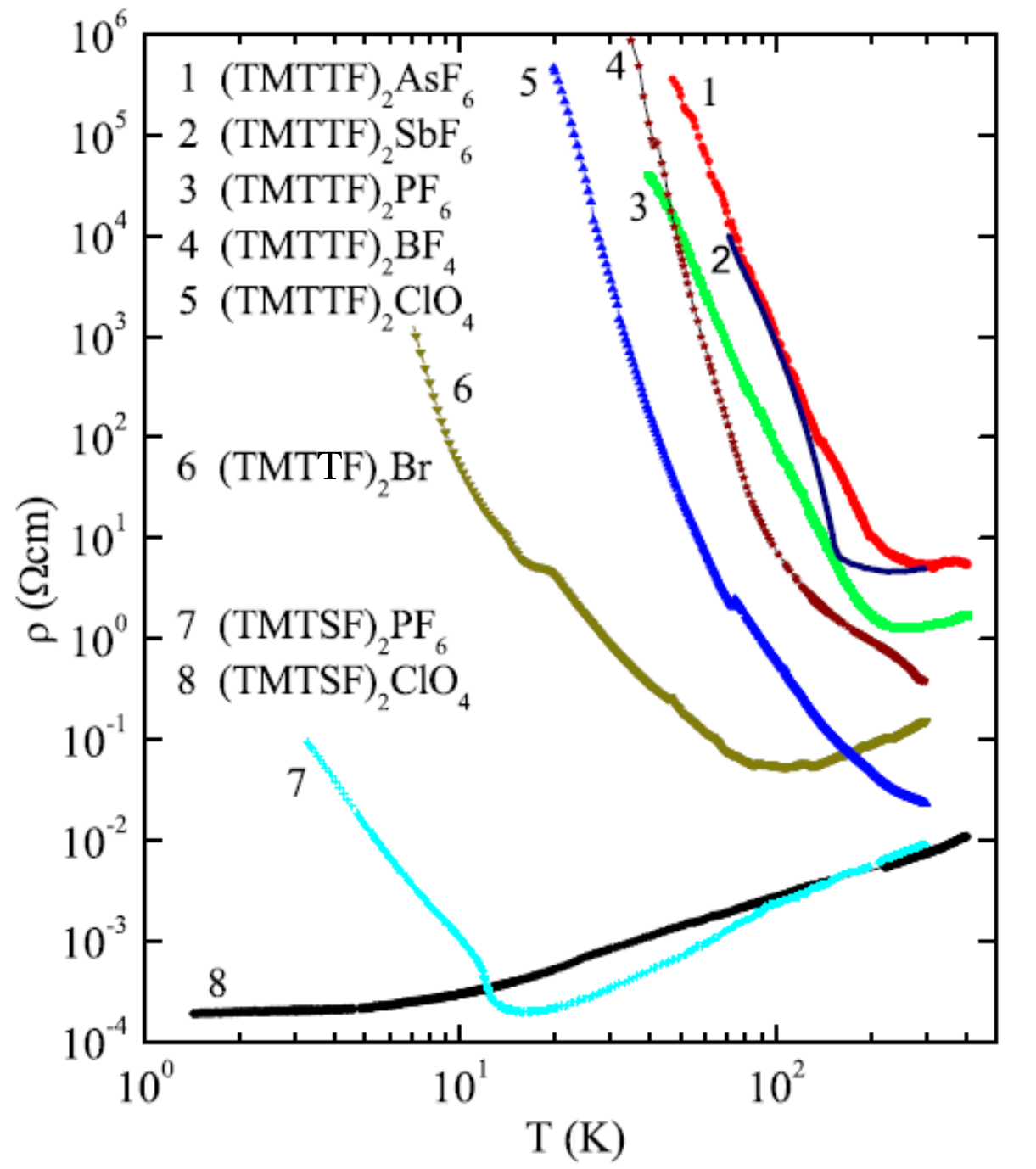}
\end{center}
\caption{\small Resistivity versus temperature for several
(TMT\emph{C}F)$_2$X salts. Upon cooling, a continuous charge
localization, marked by a broad  minimum around $T_{\rho}$ = 200 --
250\,K, is observed for the (TMTTF)$_2$AsF$_6$ (curve 1) and
(TMTTF)$_2$PF$_6$ (curve 3)
 salts before the occurrence of the
CO phase transition at much lower $T_{CO}$. For the (TMTTF)$_2$SbF$_6$ salt (curve 2), a
CO transition accompanied by a MI transition takes place at
$T_{CO,MI}$ $\simeq$ 154\,K. In (TMTTF)$_2$Br (curve 6) one has $T_{\rho} \simeq$ 100\,K before the occurrence of the AFM ground state at $T_N$ = 13\,K. (TMTSF)$_2$PF$_6$ (curve 7) remains metallic until the 12\,K SDW M-I transition.  The (TMTSF)$_2$ClO$_4$ salt (curve 8)
remains metallic down to $T_c$ = 1.2\,K, temperature below which the resistance
vanishes and superconductivity takes place. Adapted from
Ref.\,\cite{Dressel2007g7}.}\label{Resistivity-TM}
\end{figure}

By cooling (TMTTF)$_2$PF$_6$ under ambient pressure, a progressive
Mott-Hubbard charge localization is observed around $T_{\rho}$
$\simeq$ 240\,K (see curve 3 in Fig.\,\ref{Resistivity-TM}).
Upon further cooling, a CO phase transition coinciding with the
onset of a ferroelectric phase takes place at $T_{CO}$ $\simeq$
65\,K. Cooling down to lower temperatures, a SP gound state
accompanied by a stack tetramerization occurs at $T_{SP}$ $\simeq$
17\,K. By applying pressure, the CO phase and the SP ground state
are suppressed.  The SP ground state transforms into an
antiferromagnetic (AFM) one, similar to the one found in
(TMTTF)$_2$Br salt  at ambient pressure. Applying additional
pressure ($\sim$ 15\,kbar), $T_{\rho}$  vanishes showing that the
Mott-Hubbard  localized (loc) state is suppressed and that the
compound becomes metallic on a large temperature range. At low $T$,
an incommensurate spin-density wave (SDW) ground state, coinciding
with a MI transition,  replaces the commensurate AFM order. Under a
pressure of 52 -- 54\,kbar the SDW is then removed and
superconductivity occurs at $T$ = 1.4 -- 1.8\,K \cite{Adachi2000},
closing the full sequence of various ground states exhibited by the
generic phase diagram of salts with centro-symmetric anions.

\vspace{0.4cm}

\textbf{2. Non Centro-symmetric anions}

In the Fabre-Bechgaard salts incorporating non-centrosymmetrical anions, other structural phase transitions due to the ordering of the anions
 (AO transition), not shown in the generic phase diagram (Fig.\,\ref{Dressel}), occur. At RT, the anion cavities delimited by the methyl groups of
the TMT$C$F molecules are slightly disordered. Upon cooling, the change of the
spacing between TMT$C$F molecules is altered, implying thus a shortening
of contact distances between the anions and its vicinity. Due to gain of energy by reinforcing the short contact distances with the TMTCF the non-centrosymmetrical (tetrahedral) anions, such as ClO$_4$, ReO$_4$ and (linear) SCN anion order by choosing one of their two
possible orientations in the methyl groups cavity. These two possible orientations are related by inversion symmetry in the double potential \cite{foot1} originated from the anion surrounding. The barrier height between the two potential minima controls the kinetics of the AO transition
\cite{Pouget96}.
The AO transition breaks the inversion symmetry of the methyl group cavity. Also the associated deformation of the cavity perturbs the TMTCF stacking in the \emph{a}-\emph{b} layers which changes  drastically the electronic structure of the Fabre-Bechgaard salts with non centro-symmetric anions. For example, in (TMTSF)$_2$ReO$_4$ and BF$_4$ the AO transition drives a MI transition stabilizing a 2$k_F$ BCDW ground state.  In (TMTSF)$_2$ClO$_4$  the kinetics of AO drives
the salt to either a SDW ground state (quenched samples) or a
superconducting ground state (relaxed samples), for more details see
\cite{Ishiguro20,Crystals,Foury2013}.

\vspace{1.4cm}

\textbf{B.  The symmetry breaking CO transition}

\begin{figure}[!htb]
\begin{center}
\includegraphics[angle=0,width=0.47\textwidth]{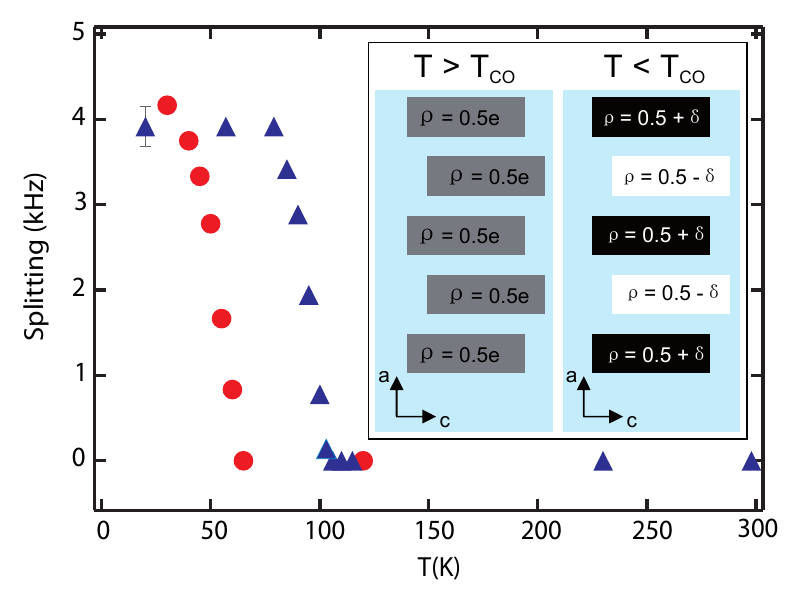}
\end{center}
\caption{\small Spectral $^{13}$C NMR line splitting versus
temperature due to two inequivalent donor molecules with unequal
electron densities in (TMTTF)$_2$X charge-transfer salts, with X =
PF$_6$ (red curve) and AsF$_6$ (blue curve). Taken from
Ref.\cite{Chow9a3}. Inset shows that above $T_{CO}$ the charge
($\rho$ = 0.5$e$) is equally distributed along the stacks. Gray
rectangles represent equally charged TMTTF molecules.  Below
$T_{CO}$ an alternated charge pattern $\pm \delta$ is found. Black
and white rectangles represent, charge rich- and charge
poor-molecules.}\label{SpectralSplitting}
\end{figure}

The AO transition of the SCN salt has a
subtle connection with the CO transition which is the main purpose of this
review. The story starts 30 years ago with the report in (TMTTF)$_2$SCN of
an AO transition accompanied by a sharp MI transition stabilizing at
$T_{CO}$ = 160\,K with formation of a superstructure with a new 4\,$k_F$ site
periodicity  \cite{Coulon82a};  a result  which was very soon
interpreted as the first evidence of a 4\,$k_F$  charge localization
(now labeled CO) triggered by the potential due to the ordering of the SCN anion \cite{Emery83,
Brazovskii85}. However, due to the interchain staggered AO process the CO does not remove all the  inversion centres of the structure. Thus the SCN salt
achieves anti-ferroelectricity below $T_{CO}$ = 160\,K. It takes a much longer time to estabilish that (TMTTF)$_2$X salts with
octahedral anions can also achieve CO where the removing of all the inversion centres leads to ferroelectricity. Two years after the work on (TMTTF)$_2$SCN, electrical
measurements evidence a similar sharp MI in the X = SbF$_6$ salt at
a critical temperature of 154\,K, which significantly shifts in the
solid solutions of SbF$_6$ with AsF$_6$ and PF$_6$
\cite{Laversanne84}. These transitions were found coinciding with a
dramatic change in the thermopower \cite{Coulon85} and a peak
divergence in the real part of the dielectric constant
\cite{Javadi88}. However, neither superstructure formation nor
significant change in the Bragg intensity was found at these
transitions, in contrast to the finding in the SCN salt, so that the
transition was labelled \emph{structureless} in the literature. The
first evidence of a symmetry breaking was found in 2000 when NMR
measurements performed by Chow \emph{et al}.\,\cite{Chow9a3}
revealed a splitting of the spectral line below the
\emph{structureless} transition, cf.\ Fig.\,\ref{SpectralSplitting}.

This spectral splitting is brought about because the TMTTF
molecules, which are equivalent above $T_{CO}$, become
differentiated below $T_{CO}$ into charge-rich alternating with
charge-poor molecules (see inset of Fig.\,\ref{SpectralSplitting} and Fig.\,\ref{COV} (a)).
This differentiation gives rise to two inequivalent hyperfine
couplings and a doubling of the spectral lines. The spectral
splitting associated with the CO is the order parameter of the
transition. Hence, based on this observation, Chow \emph{et al.}
deduced that the so-called \emph{structureless} transition in the
Fabre salts with octahedral anions  is in fact associated with a
symmetry breaking CO phase transition.


\begin{figure}
\begin{center}
\includegraphics[angle=0,width=0.47\textwidth]{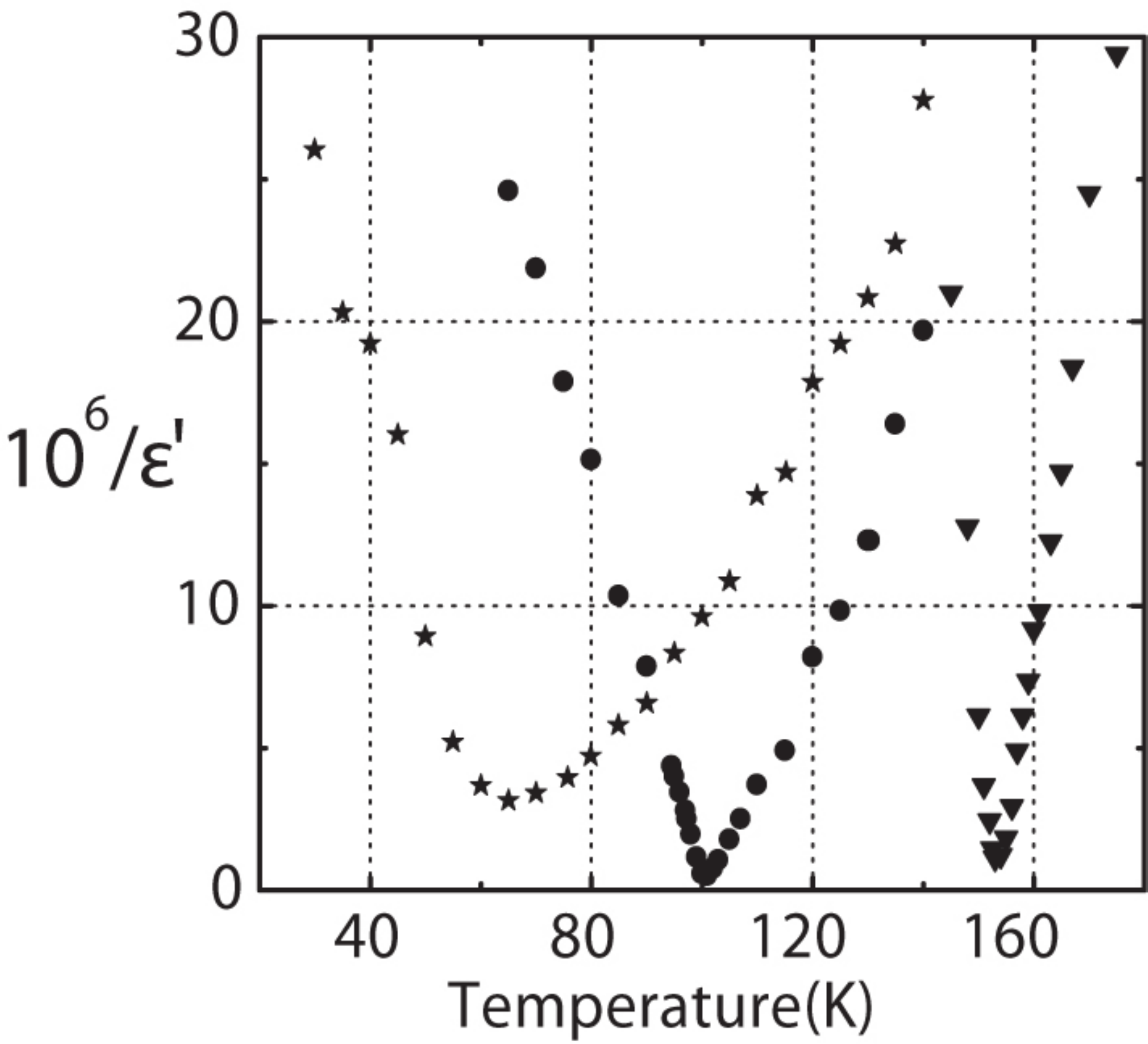}
\end{center}
\caption{\small Inverse of the real part of the dielectric constant
$\epsilon'$ of the (TMTTF)$_2$X charge-transfer salts, with X =
PF$_6$ (stars), AsF$_6$ (circles) and SbF$_6$ (triangles down),
measured at 100\,Hz. Taken from
Ref.\,\cite{Monceau2001-20}.}\label{DielectricMeasurements}
\end{figure}

Nearly at the same time, dielectric measurements
\cite{Monceau2001-20, Nad2000} were probing unambiguously the ferroelectric
 character of the transition. As a matter of fact, the first observation of a
divergence of the dielectric constant at the \emph{structureless}
transition was reported much earlier in Ref.\,\cite{Javadi88}, but the data were not interpreted in relationship with the ferroelectricity. The
features observed in the dielectric constant is due to the presence
of electric dipoles generated by the charge disproportionation along
the dimerized stacks.

Later, Dumm \emph{et al.} \cite{Dumm05-20} studied the influence of
charge disproportionation on the vibrational spectra of the X =
PF$_6$ and AsF$_6$ salts via mid-infrared optical conductivity as
a function of temperature with the light polarized along the
stacks (\emph{a}-axis) (Fig.\,\ref{MIDIR}). They found that the
intermolecular $a_g$($\nu_3$) mode, which becomes infra-red active
through electron-molecular-vibration coupling, splits below $T_{CO}$. This
feature is explained by the strong dependence of this vibronic
mode with  the degree of ionization of the TMTTF molecule
\cite{Menegheti1984}. From this study, it is estimated a
charge disproportionation ratio of 5 : 4 and 2 : 1 for X = PF$_6$
and AsF$_6$ salts, respectively.

The absence of structural effects
accompanying the CO transition was particularly puzzling as atomic
displacements, breaking the inversion symmetry, are prerequisite for
ferroelectricity to occur. It is only recently that a structural
modification was detected at the CO transition of the PF$_6$ salt using neutron
diffraction \cite{Foury2010,Pouget-PPS2012}, since it was realized
that in conventional  structural investigations the CO is rapidly
destroyed by irradiation defects induced  by the X-ray diffraction beam
\cite{Coulon-2007}. As the CO structure could not be refined, the
expected shift of the anion from the inversion centres could not be
quantified.  However, the key role of the anions, already pointed out in
earlier investigations \cite{Laversanne84},  is evidenced by the
fact that $T_{CO}$ varies significantly with the nature of the anion\cite{Pouget2007}, see Figs.\,8, 9 and 10. Note that only recently a structural refinement was able to prove that an anion
shift stabilizes
 the CO pattern in the parent salt
$\delta$-(EDT-TTF-CONMe$_2$)$_2$Br  \cite{Zorina2009}.\linebreak

\textbf{C.  Characteristic features of the CO ground state}

As Fig.\,\ref{DielectricMeasurements} demonstrates, the transition is
due to the Curie-Weiss divergence of the real part of the
dielectric constant, i.e.\ $\epsilon'$ = $A$/$\mid$ $T$
$-$$T_{CO}$$\mid$, consistent with a second-order phase transition
with a mean-field character. This shows that the regime of 1D
pretransitional fluctuations expected in 1D conductors is removed by the 3D long-range Coulomb
interactions \cite{Monceau2001-20}. Nevertheless, a careful analysis
of the dielectric constant data reveal precursor effects already at
$T \simeq$ 200\,K for the X = PF$_6$ salt \cite{Nad2000}.
Interestingly enough, below $T \simeq$ 200\,K the spin
susceptibility continues to decrease monotonically \cite{Dumm2000,Coulon82,Pouget2007}. Hence, the
development of charge correlations in the dielectric constant
together with the absence of magnetic signatures at $T_{CO}$, offer strong
evidence for the occurrence of a spin-charge separation accompanying the
CO instability.

\begin{figure}
\begin{center}
\includegraphics[angle=0,width=0.49\textwidth]{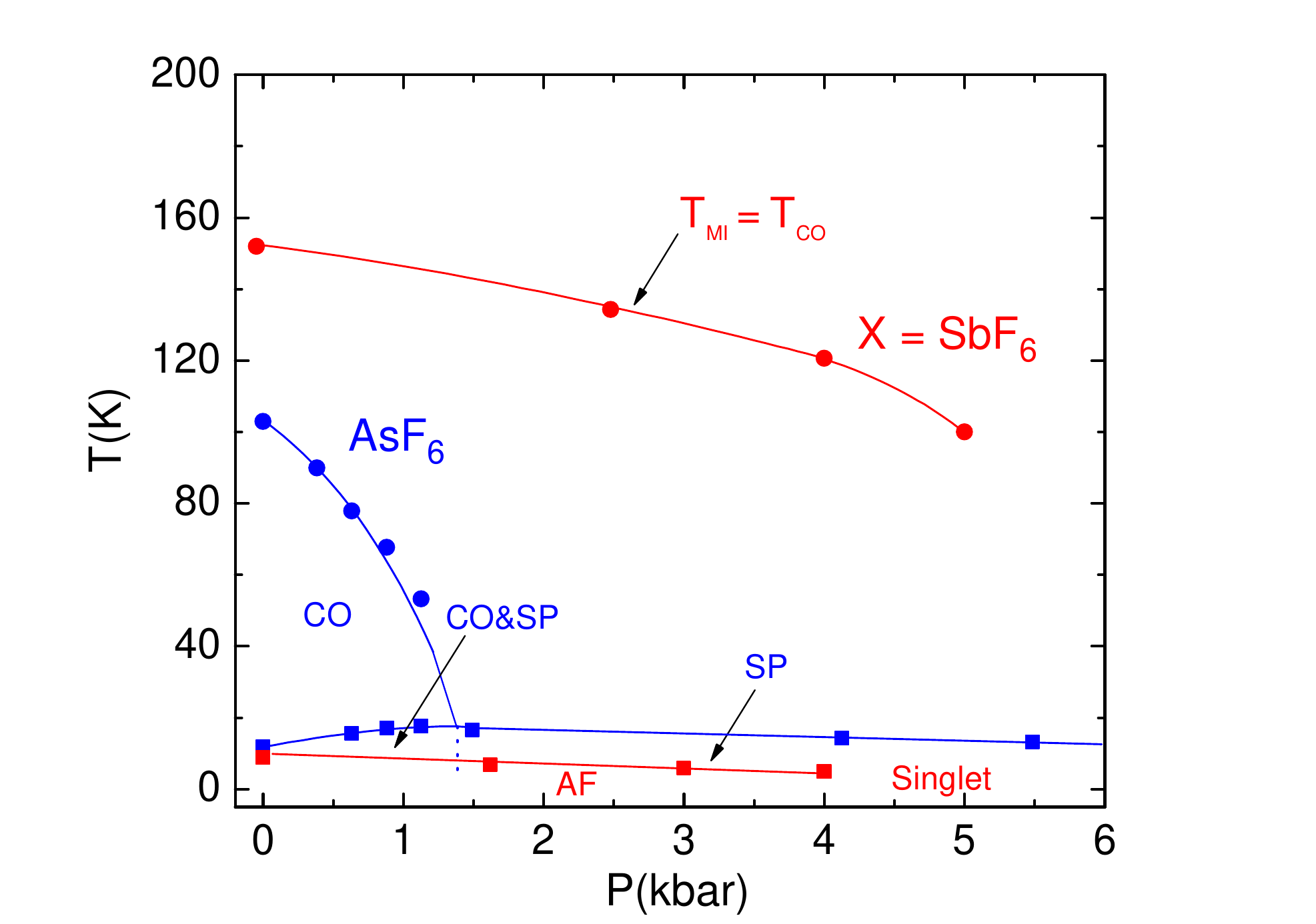}
\end{center}
\caption{\small Pressure versus temperature phase diagram for
(TMTTF)$_2$X, X = AsF$_6$ (blue) (data taken from
Ref.\,\cite{Zamborsky2002-20}) and SbF$_6$ (red) (data taken from
Ref.\,\cite{Yu2004-20}) obtained from NMR measurements under
pressure. Lines are guides to the eyes for both set of data. For X =
SbF$_6$: $T_{MI}$ = $T_{CO}$ indicates that the MI transition
temperature $T_{MI}$ coincides with the charge-ordering transition
temperature $T_{CO}$. $T_{CO}$ decreases under increasing  pressure.
Above 5\,kbar, no signature of  CO is observed.
Above 4\,kbar the AF ordering is suppressed giving rise to  local magnetic singlets or to a spin-liquid behavior. For X = AsF$_6$: CO $\&$ SP indicate the coexistence of
charge-ordered and Spin-Peierls phases. The dashed blue line separates
this coexistence region from a pure SP ground state.}\label{NMRunderP}
\end{figure}

The effect of pressure on the CO transition was studied in detail
via NMR measurements  \cite{Zamborsky2002-20,Yu2004-20}, cf.\
Fig.\,\ref{NMRunderP}. It was observed that by applying pressure,
$T_{CO}$ decreases dramatically. For example, for the X = AsF$_6$
salt (data in blue in Fig.\,\ref{NMRunderP}), a pressure of about
1.5\,kbar suppresses the CO phase.
 As can be seen from Fig.\,\ref{NMRunderP},
at lower pressures ($P$ $<$ 1.5\,kbar) both  SP and CO phases
coexist in the AsF$_6$ salt. More precisely, this figure shows that
the SP transition temperature increases when the CO is depressed.
This is a general behavior \cite{Pouget2012,Crystals} showing that
there is a repulsive coupling between the SP and CO components of
the  4$k_F$ CDW-SP ground state  (Fig.\,1e). This is confirmed by
recent accurate dielectric measurements showing a decrease of the
gap of charge at the SP transition in the PF$_6$ salt
\cite{Langlois}. For the X = SbF$_6$ salt (data in red in
Fig.\,\ref{NMRunderP}), a pressure of roughly 5\,kbar is necessary
to abruptly suppress the CO phase. At low temperatures, pressure
destroys the AF ordering giving rise to the formation of local
SP-like magnetic singlets \cite{Yu2004-20} together with
spin-liquid like fluctuations \cite{Iwase2011}. Contrary to the
expectation of the generic phase diagram (Fig.\,\ref{Dressel}), a
long-range SP ground state is not observed in pressurized
(TMTTF)$_2$SbF$_6$ as it is the case in the PF$_6$ and AsF$_6$ salts
at ambient pressure.  At higher pressure, however, the AF ground state
is recovered in the SbF$_6$ salt \cite{Iwase2011}. All these results
demonstrate that some caution must be taken in the use of the
generic phase diagram    which ignores the specificity of the
anions. In fact, anions influence the phase diagram by their
kinetics of ordering (case of (TMTSF)$_2$ClO$_4$) or by their
blockade in the methyl group cavities squeezed by the pressure. The
freezing of the orientation and translation degrees of freedom of
the anions prevents  the development of the structural counterpart
of the SP order and kills the CO in pressurized (TMTTF)$_2$SbF$_6$
(this general statement is developed in Ref.\,\cite{Crystals}).

\begin{figure}
\begin{center}
\includegraphics[angle=0,width=0.47\textwidth]{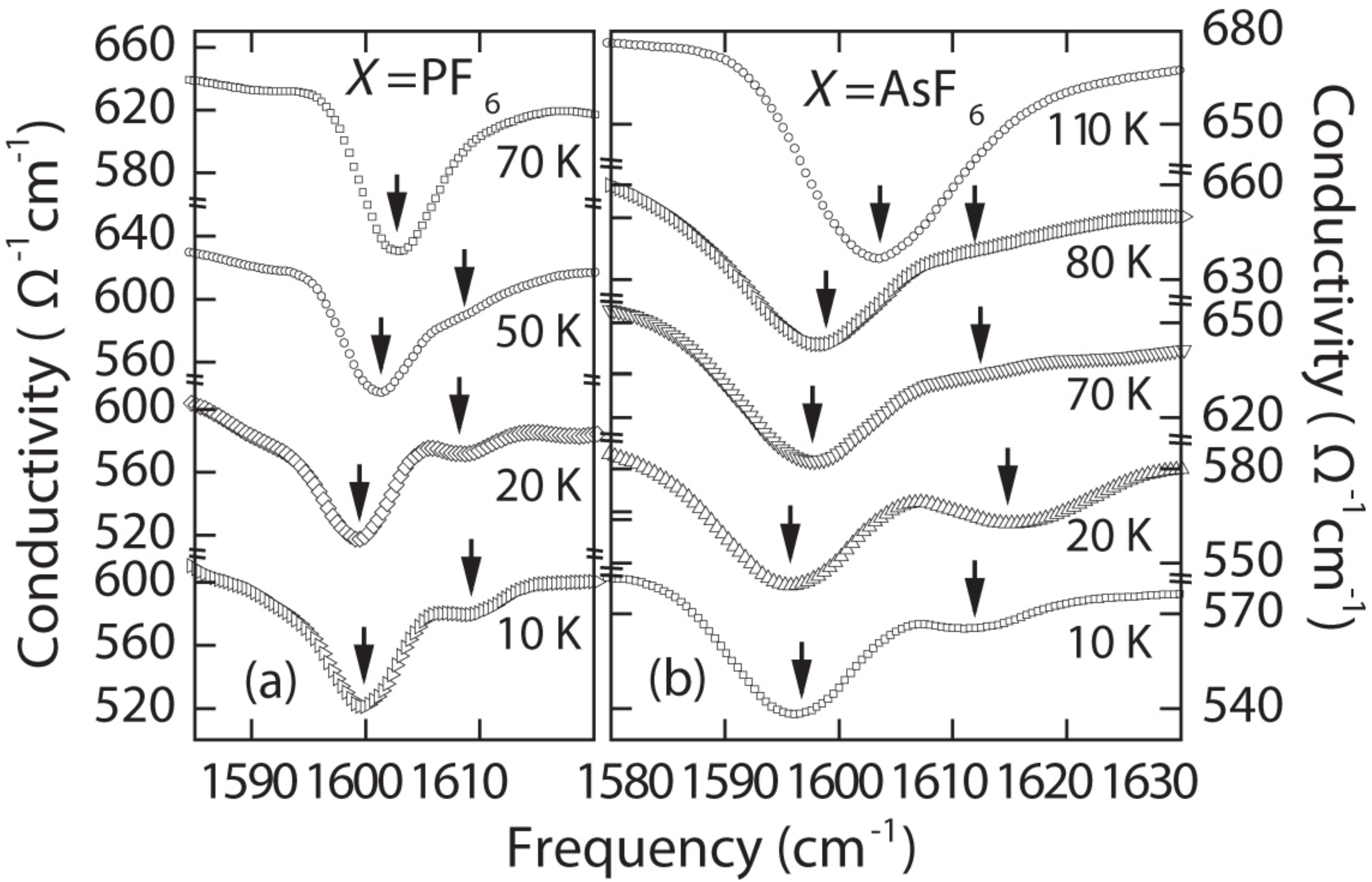}
\end{center}
\caption{\small Mid-infrared conductivity for light polarized
parallel to the stacks (\emph{a}-axis) of (TMTTF)$_2$X with (a) X
= PF$_6$ and (b) X = AsF$_6$ at various temperatures. Below
$T_{CO}$ the electron-molecular vibronic mode $a_g$($\nu_3$)
splits, as indicated by arrows, indicating the existence of two
differently charged TMTTF molecules. Taken from
Ref.\,\cite{Dumm05-20}.}\label{MIDIR}
\end{figure}

Note (Fig.\,\ref{DielectricMeasurements})
that a rather smooth divergence of the dielectric constant occurs
for the X = PF$_6$ salt at $T_{CO}$, which contrasts with
the pronounced singularity observed for the AsF$_6$ and SbF$_6$ salts.
For (TMTTF)$_2$PF$_6$, it has been found that with decreasing frequency, the maximum
of $\epsilon'$ is enhanced and its position shifts to lower
temperatures \cite{Nad2000, Nad2006-20} while for the (TMTTF)$_2$AsF$_6$ and
(TMTTF)$_2$SbF$_6$ salts the peak position remains practically
unaffected, but the magnitude of the anomaly decreases as the
frequency is increased.
This peculiar behavior of the PF$_6$ salt is well documented in the literature. It corresponds to the response of ferroelectric relaxors, see
e.g.\,Ref.\,\cite{Cheng1998}.
Such frequency dependent dielectric response indicates that ferroelectricity is probably achieved on local domains present in the PF$_6$  salt below $T_{CO}$. This means, more generally, that CO long-range order can be easily disrupted by structural defects and disorders as previously suggested by the observation of  a drastic effect of X-ray irradiation defects on the CO.  Such defects should nucleate ferroelectric domain walls. Its consequence will be a smearing out of the collective charge response at $T_{CO}$  without decrease of the CO critical temperature, see Ref.\,\cite{Coulon-2007}.  Peculiar features of the dielectric response in the Fabre salts could be also related to kinetic effects related to the order-disorder or relaxational  nature of the pretransitional dynamics of the ferroelectric phase transition \cite{Staresinic-2006}. The dynamics of the ferroelectric transition is probably controlled by the anion shift which stabilizes the CO pattern. In this respect, the anion shift could present the same type of kinetic at the CO transition as the anion orientation at the AO transition. The order-disorder dynamics could be related to the presence inside each methyl group cavity where the anion is located, of a multi-well potential provided by the donor surrounding and whose minima reflect the tendency for each anion to form local bonds with the donors.

From the theoretical point of view, using the extended Hubbard
model coupled to the lattice, Riera and Poiblanc \cite{Riera2001}
propose that CO could be due to a cooperative effect between the
inter-site Coulomb interaction $V$ and the coupling of charges located on the TMTTF with the anions. In their model, the uniform displacement of the anions, which induce local modulation of the on-site electronic energy, are necessary to stabilize the CO phase.

Concluding this section, it is useful to recall that according to
the model proposed by Riera and Poiblanc \cite{Riera2001}, and to the arguments discussed previously, anion
displacement seems to play a crucial role in the stabilization of the CO
 pattern. However, the observation of such lattice effects accompanying the CO transition was done only recently\cite{Foury2010,Zorina2009}. Before this observation, systematic
high-resolution thermal expansion experiments carried out on
the (TMTTF)$_2$X family with X = PF$_6$, AsF$_6$ and SbF$_6$ have brought the first clear-cut evidence of the involvement of the lattice in the CO instability.
These experimental results
will be presented and discussed in Sections VI and VII. Before doing so we introduce, in Section V, the Physics deduced from thermal expansion measurements.

\vspace{0.4cm}

\section{Thermal Expansion and Thermodynamic Quantities}
Thermal expansion at constant pressure quantifies the temperature
($T$) dependence of the sample volume ($V$). Upon increase or
decrease of the temperature in the vicinity of a generic phase
transition, which might have its origin in magnetic or electronic
effects, but should be accompanied by some structural effects, the
harmonic approximation is no longer valid and the crystal expands or
contracts until it finds the volume where the total free energy is
minimized. In this sense, high-resolution measurements of the
thermal expansion coefficient can be seen as  a  powerful
thermodynamic experimental tool for detecting phase transitions,
since phase transitions of different natures  can be observed by
using this method. In particular, given the high compressibility of
molecular conductors, high-resolution thermal expansion experiments
have been shown as an appropriate tool for exploring various
intriguing physical phenomena occurring in these materials
\cite{Mariano2008,Mott,Thesis,SL,RSI,Bartosch}.

The volumetric thermal expansion coefficient of a solid  is given by:

\vspace*{0.4cm}

\begin{equation}
\beta(T) = \frac{1}{V} \Bigl(\frac{\partial V}{\partial T}\Bigl)_P
\label{te}
\end{equation}

\vspace*{0.4cm}

where,

\vspace*{0.4cm}

\begin{equation}
\beta(T) = \alpha_a(T) + \alpha_b(T) + \alpha_c(T) \label{alfa}
\end{equation}

\vspace*{0.4cm}

and $\alpha_i$ is the linear thermal expansion coefficient along
the \emph{i} = \emph{a}, \emph{b} and \emph{c} crystal directions.
Eq.\,\ref{alfa} holds for all lattice symmetries if \emph{a},
\emph{b} and \emph{c} are perpendicular to each other
\cite{Barron9f}. The linear thermal expansion coefficient at
constant pressure ($P$) reads:

\vspace*{0.4cm}

\begin{equation}
\alpha_{i} = \frac{1}{\emph{l}} \Bigl(\frac{\partial
l(T)}{\partial T}\Bigl)_P \label{linear coeficient}
\end{equation}

\vspace*{0.4cm}

where \emph{l} is the sample length in the direction of measurement. The physical quantity
described by Eq.\,\ref{linear coeficient} will be frequently used
in this paper.

The isothermal compressibility of a solid is defined as follows:

\vspace*{0.4cm}

\begin{equation}
\kappa_{T} = -\frac{1}{V} \Bigl(\frac{\partial V}{\partial
P}\Bigl)_T \label{compressibility}
\end{equation}

\vspace*{0.4cm}

In order to link  the volumetric thermal expansion
coefficient (Eq.\,\ref{te}) of a solid and the isothermal compressibility (Eq.\,\ref{compressibility}), the derivative entering in
Eq.\,\ref{te} can be  decomposed in the following way:

\vspace*{0.4cm}

\begin{equation}
\beta(T) = -\frac{1}{V} \frac{\partial V}{\partial P}\Bigl|_{T}
\cdot \frac{\partial P}{\partial
T}\Bigl|_{V}=\kappa_{T}\frac{\partial P}{\partial T}\Bigl|_{V}
\label{beta1}
\end{equation}

\vspace*{0.4cm}

The volumetric thermal expansion coefficient can still be linked
with the entropy (\emph{S}) of the system. To this end, it is
useful to work with the Helmholtz free energy $F(V, T)$, defined
as follows:

\vspace*{0.4cm}

\begin{equation}
F = U-TS \label{Helmholtz}
\end{equation}

\vspace*{0.4cm}

where $U$ refers to the internal energy of the system. Making
the partial derivatives of Eq.\,\ref{Helmholtz}, one obtains:

\vspace*{0.4cm}

\begin{equation}
\frac{\partial F}{\partial T}\Bigl|_{V} = -S \label{entropy}
\end{equation}

\vspace*{0.4cm}

\begin{equation}
\frac{\partial F}{\partial V}\Bigl|_{T} = -P \label{pressure}
\end{equation}

\vspace*{0.4cm}

Including Eq.\,\ref{pressure} into Eq.\,\ref{beta1} results in:

\vspace*{0.4cm}

\begin{equation}
\beta = -\frac{1}{V}\frac{\partial V}{\partial P}\Bigl|_{T}
\Bigl[\frac{\partial}{\partial T} \cdot \Bigl(-\frac{\partial
F}{\partial V}\Bigl)\Bigl|_{T}\Bigl]\Bigl|_{V} \label{beta2}
\end{equation}

\vspace*{0.4cm}

Interchanging the derivative in the above equation and substituting
Eqs.\,\ref{compressibility} and \ref{entropy}, results in:

\vspace*{0.4cm}

\begin{equation}
\beta(T) = - \kappa_T \frac{\partial^2 F}{\partial T \partial V} =
\kappa_T  \frac{\partial S}{\partial V}\Bigl|_{T} \label{beta4}
\end{equation}

\vspace*{0.4cm}

Eq.\,\ref{beta4} shows the direct connection of the volumetric
thermal expansion coefficient to the volume dependence of the
entropy. In the following, the volumetric thermal expansion
coefficient will be related to the specific heat, which is defined
as the amount of heat $Q$ necessary to increase the temperature of
the sample, as follows:

\vspace*{0.4cm}

\begin{equation}
C(T) = \frac{\Delta Q}{\Delta T}  \label{specific heat definition}
\end{equation}

\vspace*{0.4cm}

The specific heat at
constant volume is defined from the Helmholtz free energy by:

\vspace*{0.4cm}

\begin{equation}
C_V(T) = -T \frac{\partial^2 F}{\partial T^2}\Bigl|_{V} = T
\frac{\partial S}{\partial T}\Bigl|_{V} \label{specific heat}
\end{equation}

\vspace*{0.4cm}

Eq.\,\ref{beta4}  can be rewritten as follows:

\vspace*{0.4cm}

\begin{equation}
\beta(T) = - \kappa_T \frac{\partial S}{\partial T}\Bigl|_{V}
\cdot \frac{\partial T}{\partial V}\Bigl|_{S}\label{beta5}
\end{equation}

\vspace*{0.4cm}

Substituting the last part of Eq.\,\ref{specific heat} into
Eq.\,\ref{beta5} and using the identity $(V/T) \cdot \partial T / \partial V =
\partial \ln T / \partial \ln V$ the desired relation between $C_V(T)$ and $\beta(T)$ is thus obtained:

\vspace*{0.4cm}

\begin{equation}
\beta(T) = -\kappa_T \cdot C_V(T) \cdot \frac{1}{V} \cdot
\frac{\partial \ln T}{\partial \ln V}\Bigl|_{S}\label{beta6}
\end{equation}

\vspace*{0.4cm}

By defining the new quantity

\begin{equation}
\Gamma = -\frac{\partial\ln T}{\partial\ln
V}\Bigl|_{S}\label{gamma}
\end{equation}

\vspace*{0.4cm}

equation \ref{beta6} becomes

\vspace*{0.4cm}

\begin{equation}
\beta(T) = \Gamma_{eff} \cdot \frac{\kappa_T}{V_{mol}} \cdot
C_V(T) \label{Grueneisen}
\end{equation}


The latter equation is called Gr\"uneisen-Relation
\cite{Grueneisen1908}, where $V_{mol}$ stands for the molar volume
and $\Gamma_{eff}$ is the effective Gr\"uneisen parameter. In
general, $\Gamma_{eff}$$\cdot$$\kappa_T$/$V_{mol}$ is weakly
temperature dependent.

In the frame of the Debye model, the phononic specific heat reads:
\begin{equation}
C_V = 9 N k_B \biggl(\frac{T}{\Theta}\biggl)^3 \int \limits_0^{x} \! {\frac{x^4 e^x}{(e^x - 1)^2}} dx ,
\label{Debye}
\end{equation}
where $x$ = $\Theta/T$,  $\Theta$ stands for the Debye temperature, $N$ is the number of atoms taken into account and $k_B$ is the Boltzmann constant. To fit the thermal expansion data shown in Fig.\,\ref{TE-PF6} (see Section \ref{Thermal Expansion on TMTTF}), we combined Eqs.\,\ref{Grueneisen} and \ref{Debye}, as follows:
\begin{equation}
\beta(T) = \Gamma_{eff}  \frac{\kappa_T}{V_{mol}}
9  N \cdot k_B  \biggl(\frac{T}{\Theta}\biggl)^3 \int \limits_0^{x} \! {\frac{x^4 e^x}{(e^x - 1)^2}} dx \label{Debye-Fits}
\end{equation}
In such fits (Fig.\,\ref{TE-PF6}), the pre-factor ($\Gamma_{eff}   \frac{\kappa_T}{V_{mol}}
9   N   k_B$) of the integral in Eq.\,\ref{Debye-Fits} and $\Theta$ were considered as fit parameters.

The lattice (or phononic) Gr\"uneisen parameter $\Gamma_{pho}$ is given by:

\vspace*{0.4cm}

\begin{equation}
\Gamma_{pho} = -\frac{d\ln \Theta_D}{d\ln V}  ,
\label{LatticeGrueneisenParameter}
\end{equation}
\vspace*{0.4cm}

According to
Eq.\,\ref{LatticeGrueneisenParameter}, the bigger the lattice
Gr\"uneisen parameter, the higher the volume dependence of the
vibration modes of the lattice. Strictly speaking, the lattice
Gr\"uneisen parameter is a measure of the volume dependence of the
anharmonicity of the lattice vibrations, which in turn is
responsible for the lattice contribution to the thermal expansion
in a solid. If the vibrational free energy, entropy, specific heat
and thermal expansion are the sum of contributions $f_i$,
$s_i$, $c_i$ and $\alpha_i$ due to independent vibration modes of
frequency $\omega_i(V)$, respectively,  it is convenient to define the Gr\"uneisen
parameter  of the phonon mode \emph{i} in the following way:

\vspace*{0.4cm}

\begin{equation}
\Gamma_{i} = -\frac{d\ln \omega_i}{d\ln V}\label{Grueneisen modes}
\end{equation}

\vspace*{0.4cm}

Thus, according to Eq.\,\ref{Grueneisen modes}, vibration modes
whose frequency, $\omega_i$, decreases or \emph{softens} as the
volume of the solid decreases will lead to a negative Gr\"uneisen
parameter and, from Eq.\,\ref{Grueneisen}, these modes will
be responsible for a negative contribution to the overall thermal
expansion of the material \cite{NTE}.

More generally, in addition to the phonon contribution
at the thermal expansion of a material, other contributions, whose
origin might be electronic or magnetic, have to be included. This is especially the case at low temperatures, where
such contributions may dominate the thermodynamic properties
\cite{Barron9f}. Hence, the total volumetric thermal expansion
coefficient can be generally expressed as:

\vspace*{0.2cm}

\begin{equation}
\beta  =  \beta_{ph}+ \beta_{el} + \beta_{mag} =
\frac{\kappa_T}{V_{mol}}(\Gamma_{ph}C_{ph} +  \Gamma_{el}C_{el} +
\Gamma_{mag}C_{mag}),\label{Grueneisen electronic and magnetic}
\end{equation}

\vspace*{0.2cm}

where $\beta_{ph}$ ($C_{ph}$), $\beta_{el}$ ($C_{el}$) and
$\beta_{mag}$ ($C_{mag}$) refer to the phononic, electronic and
magnetic contributions to $\beta$ ($C$), respectively, while
$\Gamma_{ph}$, $\Gamma_{el}$ and $\Gamma_{mag}$ are the respective
Gr\"uneisen parameters.

If the Gr\"uneisen parameter, which measures
the volume dependence of a characteristic temperature, is constant and if in the temperature range of interest only one of  its
contributions is predominant, then it is expected that $C_V(T)$ and $\beta(T)$  will have the same temperature dependence. For this reason thermal expansion measurements provide a good test of the mean-field (jump of $C_V$($T$)) or critical ($\lambda$-type divergence of $C_V$($T$)) character of a second-order phase transition. In the latter case critical exponents can be obtained via thermal expansion measurements \cite{Bartosch}.

Interestingly, from the volume dependence of the critical temperature of the superconducting
transition in $\kappa$-(BEDT-TTF)$_2$Cu(NCS)$_2$ one obtains, by analogy with Eq.\,(17), $\Gamma \approx $ 40 \cite{Jens2000-9c}, a value
which roughly exceeds by a factor of twenty $\Gamma$ obtained for
ordinary superconductors such as in Pb  ($\Gamma$ = 2.4
\cite{Gladstone1969-9c}) or which is even much larger than $\Gamma$ obtained for the
layered cuprate YBa$_2$Cu$_3$O$_7$  ($\Gamma$ = (0.36 $\sim$ 0.6)
\cite{Meingast1991-9c}). These findings reveal the strong sensitivity
of the superconductivity  to lattice parameters in the family of
$\kappa$-(BEDT-TTF)$_2$X, as discussed in more detail in
Ref.\,\cite{Lang2004-9c}.

Very recently it has been reported \cite{Foury2013} that in the Bechgaard salt (TMTSF)$_2$PF$_6$  the phononic Gr\"uneisen parameter $\Gamma$, defined by expression \ref{Grueneisen modes}, is dominated by the anion rotation ($\Gamma$(PF$_6$) $\approx$ 28). As this finding remains true for the Fabre salts with octahedral anions, lattice expansion data of section VI should reveal the electronic instabilities on the TMTTF stack via the modification of the anion vibrational spectra (generally caused by a modification of the volume and shape of the methyl group cavity where the anion is located or by a modification of the linkage of the anion with the neighboring organic molecules).

\section{Thermal Expansion Measurements on (TMTTF)$_2$X
salts}\label{Thermal Expansion on TMTTF}
The focus of this Section is the direct
observation of structural changes
associated with the CO via high-resolution thermal expansion
experiments.
Thermal expansion measurements were carried
out along three orthogonal axes, namely along the \emph{a-},
\emph{b'}- and \emph{c$^*$}-axes. The \emph{a}-axis is along the
stacks, the \emph{b'}-axis is perpendicular to the \emph{a}-axis in
the \emph{a}-\emph{b} plane and \emph{c$^*$}-axis is perpendicular
to the \emph{a}-\emph{b} (\emph{a}-\emph{b'}) plane. The
crystals of  (TMTTF)$_2$X (X = PF$_6$, AsF$_6$ and SbF$_6$)
have the following shape: the long crystal axis is parallel to the
\emph{a}-axis, the intermediate axis is parallel to the \emph{b'}-axis
and the short one is parallel to the \emph{c$^*$}-axis.
As the crystal structure is triclinic, the three orthogonal directions  previously defined are not the eigen-directions of the thermal expansion tensor. Earlier measurements of the eigen-components  of this tensor using less accurate neutron diffraction methods, have shown, both for (TMTSF)$_2$PF$_6$  \cite{Gallois86}  and  (TMTTF)$_2$PF$_6$ \cite{Gallois20g5,Granier1988}, that the eigen-directions change considerably in temperature and thus deviate significantly from the orthogonal set used in the present study.

\subsection{ Overall thermal  behavior of the lattice expansion coefficient}
\begin{figure}
\begin{center}
\includegraphics[angle=0,width=0.505\textwidth]{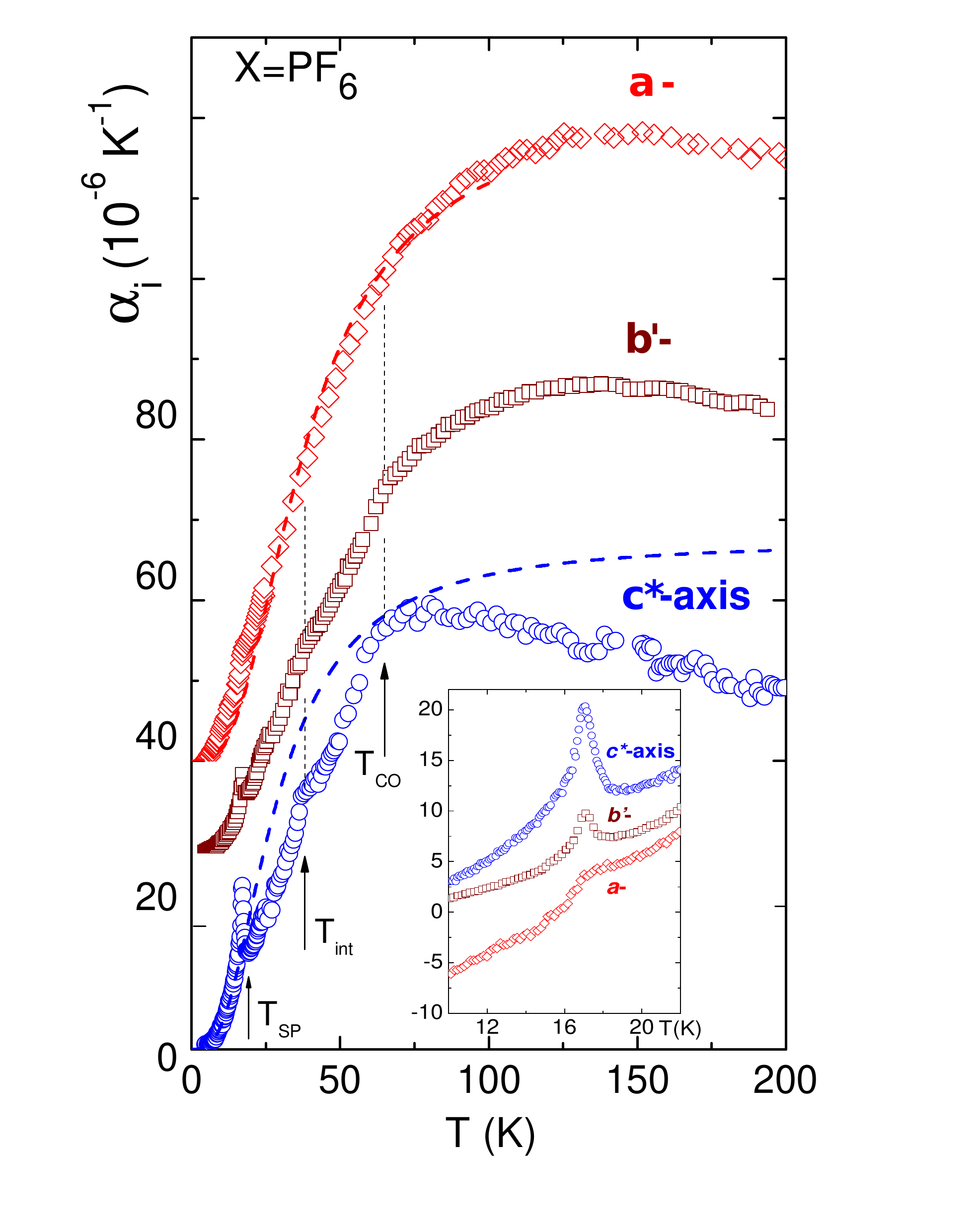}
\end{center}
\caption{\small Uniaxial thermal expansion coefficient versus $T$
along the \emph{a}-, \emph{b$'$}- and \emph{c$^*$}-axis for the
(TMTTF)$_2$PF$_6$ salt. Red dashed line is a Debye fitting up to 100\,K
along the \emph{a}-axis. Data are shifted for clarity. Blue dashed curve is a Debye fitting for the data along the \emph{c$^*$}-axis, cf.\, discussion in the main text.  Inset: zoom
of the low temperature data. Arrows indicate the charge-ordering
($T_{CO}$), the intermediary ($T_{int}$) and the Spin-Peierls
($T_{SP}$) transition temperatures. Adapted from Refs.\,\cite{Mariano2008,Thesis}.}\label{TE-PF6}
\end{figure}

\begin{figure}
\begin{center}
\includegraphics[angle=0,width=0.48\textwidth]{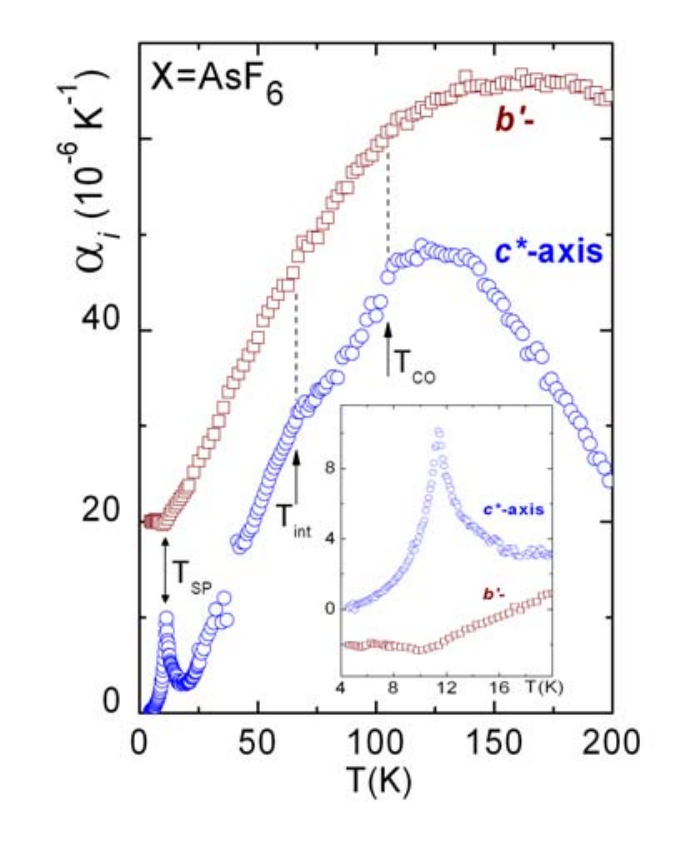}
\end{center}
\caption{\small Uniaxial thermal expansion coefficient versus $T$
along the \emph{b$'$}- and \emph{c$^*$}-axis for the
(TMTTF)$_2$AsF$_6$ salt. Data are shifted for clarity. Inset: blow
up of the low-temperature data. Arrows indicate the
charge-ordering ($T_{CO}$), the intermediary ($T_{int}$) and the
Spin-Peierls ($T_{SP}$) transition temperatures, respectively.
Missing data in $\alpha_{c^*}$ in the $T$ window 39 - 41\,K is due
to high noise in this temperature window. Taken from Ref.\,\cite{Mariano2008,Thesis}.}\label{TE-AsF6}
\end{figure}

\begin{figure}
\begin{center}
\includegraphics[angle=0,width=0.49\textwidth]{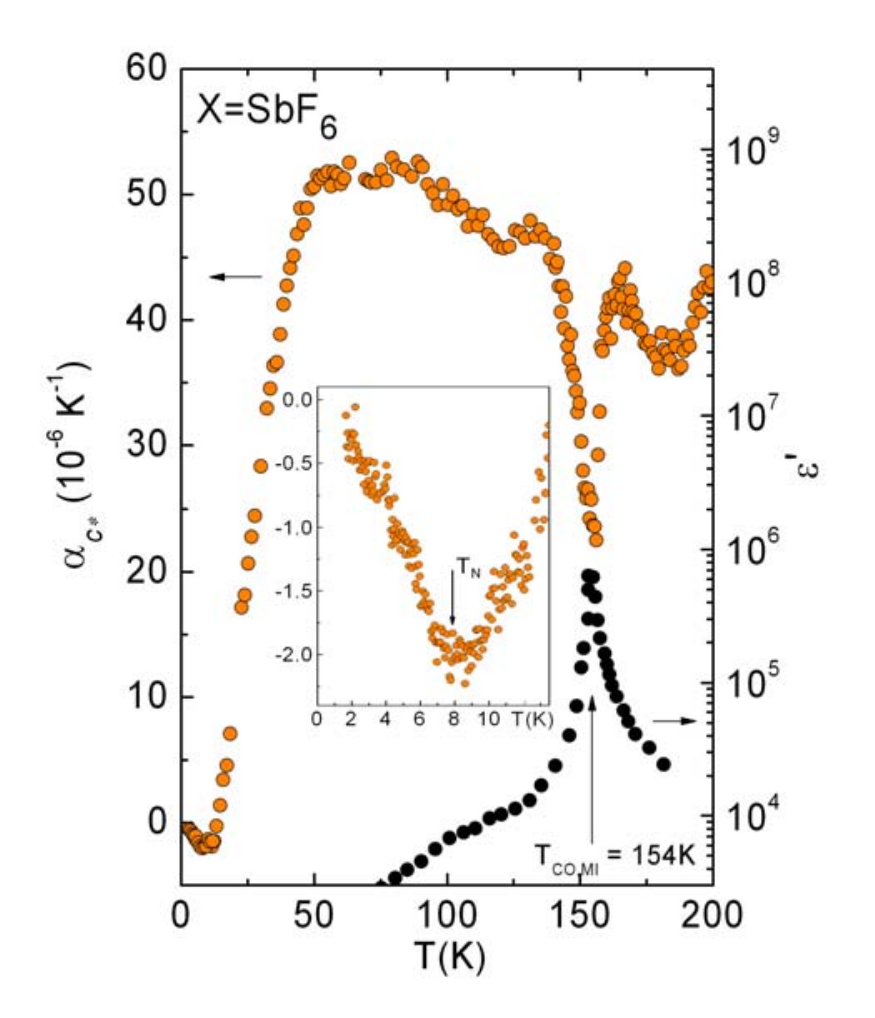}
\end{center}
\caption{\small Left scale: uniaxial thermal expansion coefficient
versus $T$ along the \emph{c$^*$}-axis for the (TMTTF)$_2$SbF$_6$
salt. Right scale: real part of the dielectric constant
$\epsilon'$ (data taken from Ref.\,\cite{Nad2006-20}) plotted on
the same $T$ scale as $\alpha_{c^*}$. $T_{CO,MI}$ indicates the
coexistence of CO and MI transitions. Inset shows details of the
tiny anomaly in $\alpha_{c^*}$ due to the antiferromagnetic
ordering at $T_N$ $\simeq$ 8\,K. Taken from Ref.\,\cite{Thesis}.}\label{SbF6}
\end{figure}

Figures \ref{TE-PF6}, \ref{TE-AsF6} and \ref{SbF6} show the uniaxial expansivity for the X = PF$_6$, AsF$_6$ and SbF$_6$ salts, respectively. As can be seen from Figs.\,\ref{TE-PF6}
and \ref{TE-AsF6}, the data show that the lattice expansivity is anisotropic at high temperature and quite isotropic at low temperature, let say below 100\,K. The largest expansivity is along the \emph{a} (stack) direction and the smallest one is in the \emph{c$^*$} direction, along which planes of TMTTF molecules alternate with planes of  counter anions X. An intermediate lattice expansivity is measured along the interstack \emph{b'} direction. The anisotropy is due to the fact that the rate of increase of the expansion coefficient upon heating saturates first along \emph{c$^*$}, then along \emph{b'}  and finally along \emph{a}, before becoming negative at higher temperature for all the directions.
For the X = PF$_6$ (Fig.\,\ref{TE-PF6}) and X = AsF$_6$ (Fig.\,\ref{TE-AsF6})
salts, an anomalous thermal dependence is observed along the \emph{c$^*$}-axis for $T
> T_{CO}$. The data reveal a striking negative slope, $d\alpha_{c^*}(T)$/$dT$ $<$
0, starting from about the CO transition temperature $T_{CO}$. Less pronounced
effects are observed along the \emph{a}- and \emph{b'}-axes.  A negative slope, $d\alpha_{c^*}(T)$/$dT$ $<$ 0, is also observed in the SbF$_6$ salt (Fig.\,14) above about 75\,K, well below $T_{CO}$.   At this point, it is interesting to remark that the negative slope regime is not observed for the uniaxial expansivity measured along \emph{c$^*$} in (TMTTF)$_2$Br, cf.\,Ref.\,\cite{Dani}. In addition, this lattice expansivity  is one order of magnitude smaller in the Br salt than in the PF$_6$,  AsF$_6$, and SbF$_6$ salts. Thus the unusual features previously described are probably due to the rotational degrees of freedom of the octohedral anions which do not exist in the Br salt.

As expected from these unusual features, attempts to fit the data employing a Debye-like behavior along the three axes were unsuccessful.  Along the chain-axis up to $\sim$ 100\,K (red dashed line in Fig.\ref{TE-PF6}) a Debye dependence fits (Eq.\,\ref{Debye-Fits}) relatively well the data, with a
Debye temperature $\Theta_D$ of 158\,K. We observed, however, that by changing
the temperature range of the fits, $\Theta_D$ varies markedly. For
example, a fitting up to 24\,K along the \emph{a}-axis results in
$\Theta_D$ = 102\,K. However, due to the anomalous lattice effects,
depicted in Figs.\,\ref{TE-PF6} and \ref{TE-AsF6} along the
\emph{b'}- and \emph{c$^*$}-axes, the Debye model fails to describe
the data, see for example the blue dashed line in Fig.\,12.\linebreak

\subsection{Lattice anomalies at the CO transition}
Figures 12, 13 and 14 show that the uniaxial thermal expansion exhibits an anomaly at the CO transition of PF$_6$ ($T_{CO} \simeq$  65\,K), AsF$_6$ ($T_{CO} \simeq$  100\,K) and SbF$_6$ ($T_{CO} \simeq$   154\,K) salts, respectively.
The strongest anomaly is observed for measurements performed   along the \emph{c$^*$}-axis. Interestingly, along the chain direction
(\emph{a}-axis)\cite{foot3}, almost no effects are observed at the CO transition. The
anomalous behavior along the \emph{c$^*$} direction, where the
stacks are separated by anions, provides strong evidence
that the anions play a crucial role in the stabilization of
the CO phase.
For the X
= PF$_6$ (Fig.\,\ref{TE-PF6}) and X = AsF$_6$ (Fig.\,\ref{TE-AsF6})
salts, it is observed upon cooling,
 a rapid decrease of $\alpha_{c^*}$ at about $T_{CO}$. This is
indicative of a broadened step-like anomaly, i.e.\ a mean-field-like
transition, with $\Delta\alpha_{c^*}\mid_{T_{CO}}$ = $\alpha (T
\rightarrow T_{CO}^-)$ $-$ $\alpha (T \rightarrow T_{CO}^+)$ $<$ 0. Indeed,
 a broad anomaly is expected in the PF$_6$ salt due to the smooth divergence of the dielectric constant at $T_{CO} $   (see Figs.\,9 and 16), but not in the case of the AsF$_6$ salt where the dielectric constant diverges at $T_{CO} $   (Fig.\,9).
$\Delta\alpha_{c^*}\mid_{T_{CO}}$ enters in the Ehrenfest relation for
second-order phase transitions:

\vspace*{0.4cm}

\begin{equation}
\left(\frac{dT_c}{dP_i}\right)_{P_i \rightarrow 0} =  V_{mol} \cdot
T_c \cdot \frac{\Delta\alpha_i}{\Delta C}\label{Ehrenfest}
\end{equation}

\vspace*{0.4cm}

where $\Delta\alpha_i$ and $\Delta C$ refer to the thermal expansion
and specific heat changes at the transition temperature ($T_c$),
respectively. In expression (24), the index \emph{i} refers to the crystallographic
direction, along which pressure is applied. Strictly speaking, the
Ehrenfest relation is valid only for mean-field-like phase
transition, where $\Delta\alpha$ and $\Delta C$ present step-like
behavior. According to this relation, the sign of the
volumetric thermal expansion coefficient jump, i.e.\ $\Delta\beta$ =
$\Sigma_i\Delta\alpha_i$, defines the pressure dependence of the
corresponding second-order phase transition. Hence, the negative
jump anomaly in $\alpha_i(T)$ at $T_{CO}$ is consistent with
$dT_{CO}$/$dP$ $<$ 0, which agrees with NMR data under hydrostatic
pressure \cite{Zamborsky2002-20}.\linebreak

\begin{figure}
\begin{center}
\includegraphics[angle=0,width=0.48\textwidth]{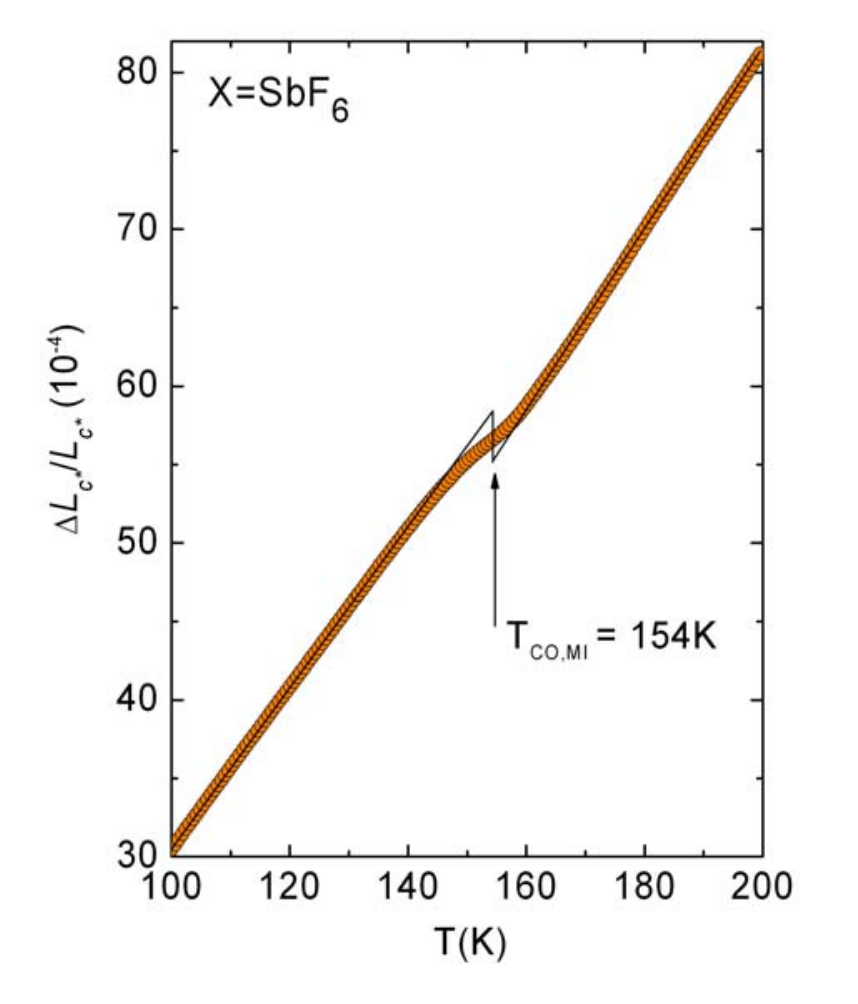}
\end{center}
\caption{\small Relative length changes versus $T$ along the
\emph{c$^*$}-axis for the (TMTTF)$_2$SbF$_6$ salt. Solid lines are
used to magnify the change of the lattice parameter at
$T_{CO,MI}$ = 154\,K. Taken from Ref.\,\cite{Thesis}.}\label{SbF6_DeltaLbyL}
\end{figure}

This behavior should be contrasted with the
huge negative $\lambda$-type anomaly observed at
$T_{CO,MI}$ = 154\,K (Fig.\,\ref{SbF6}) for the X = SbF$_6$ salt. This
anomaly coincides nicely with the peak in $\epsilon'$. Its shape is quite distinct
from the broadened step anomaly observed at the CO transition in the X = PF$_6$, AsF$_6$ salts. Interestingly, the anomaly at $T_{CO,MI}$ in X = SbF$_6$
exhibits the same shape as the anomaly observed at the MI Mott-Hubbard transition in
$\kappa$-D8-Br \cite{Mott}.
This agrees also with the fact that in the X = SbF$_6$ salt, the charge
localization does not occur gradually as in the X = PF$_6$ and
AsF$_6$ salts, but abruptly at the MI transition. As can be seen
from Fig.\,\ref{SbF6_DeltaLbyL}, the $\lambda$-type anomaly of
Fig.\,14 corresponds to  an abrupt change of the (001)
interlayer spacing with a relative jump of about 3.3
$\times$ 10$^{-4}$ at $T_{CO,MI}$ = 154\,K (solid line in Fig.\,15). This striking different
behavior of the lattice expansivity at the CO transition between the
various Fabre salts, already commented in Ref.\,\cite{Dani},
probably relies on the presence of screened electron-electron
repulsions above $T_{CO}$ in the SbF$_6$ salt while these
interactions begin to be unscreened below $T_{\rho}$, well above
$T_{CO}$, in the AsF$_6$ and PF$_6$  salts.
 It is also worth mentioning that the shape of the anomalies at $T_{CO}$ of the SbF$_6$ salt
bears some resemblance with the sharp negative $\lambda$-type anomaly
observed at the transition temperature of conventional ferroelectic insulators
like tri-glycine sulfate
(TGS) \cite{Imai1977} and BaTiO$_3$ \cite{Sawada1948}.

\subsection{Lattice anomalies at the SP and AF transitions}
The inset of Figs.\,12 and 13 shows that thermal expansion measurements exhibit a well-defined $\lambda$-type anomaly at the SP transition of the PF$_6$ and AsF$_6$ salts occurring at 17\,K and 11.4\,K, respectively. The anomaly is the strongest for measurements performed along the \emph{c$^*$} direction, weaker for those performed along  \emph{b'} and very weak for those performed along \emph{a}. The $\lambda$-type anomaly is positive for all directions of measurements in the PF$_6$ salt. Surprisingly it is negative  along the \emph{b'}  direction in the AsF$_6$ salt. Measurements performed with the same PF$_6$  sample show that the lattice expansion anomaly reflects  the critical divergence of the specific heat at the SP transition \cite{eu2009}.
Note that the anomalies at $T_{CO}$ and $T_{SP}$ differ markedly by their shapes.
While kinks observed at $T_{CO}$ are consistent with a mean-field-type
transition, as discussed above, the $\lambda$-type anomalies at $T_{SP}$ for both
salts  is compatible with the
presence of significant critical fluctuations, complying with
SP fluctuations observed for both compounds by X-ray
diffuse scattering studies \cite{Pouget82,Laversanne84}.
As for the CO transition more pronounced effects are observed for the uniaxial thermal expansion coefficient measured along the
\emph{c$^*$}-axis. This means that the tetramerization of the TMTTF
stacks achieved by the SP pairing  is accompanied by important elastic deformations perpendicular to
the stack direction, which could be induced by the transverse shift of  the TMTTF along their long molecular axis.  Since the TMTTF molecules form cavities
where the anions are located, the SP distortion of the TMTTF
sublattice should also cause a shift of the anions. The involvement of the anion in the SP transition is assessed by the measurement of a critical divergence of the NMR $^{75}$As relaxation rate at the SP transition of (TMTTF)$_2$AsF$_6$ \cite{Zamborszky-02}.\linebreak

The inset of Fig.\,\ref{SbF6} reveals that a negative anomalous contribution in
$\alpha_{c^*}$ shows up with an extremum around $T_N$ achieving the AF ground state in (TMTTF)$_2$SbF$_6$.
Interestingly, a similar negative anomalous contribution in $\alpha_{c*}$ also occur near $T_N$   in the Br salt\cite{Dani}. These behaviors should be contrasted with the observation  of a clear $\lambda$-type anomaly at the SDW and MI transition of (TMTSF)$_2$PF$_6$ \cite{Lang2004,Foury2013}.
Although very small, this negative
contribution indicates that $dT_N$ / $dP_{c^*}$ $<$ 0. This is in line with NMR measurements on the X = SbF$_6$ salt under hydrostatic pressure which reveal a negative pressure dependence of $T_N$ \cite{Zamborsky2002-20}, see
Fig.\,\ref{NMRunderP}. Hence, such results provide also evidence
that for the X = SbF$_6$ the variation of the \emph{c$^*$} lattice parameter should control the pressure-induced dependence of $T_N$.

\subsection{The lattice anomaly at $T_{int}$}
\begin{figure}
\begin{center}
\includegraphics[angle=0,width=0.50\textwidth]{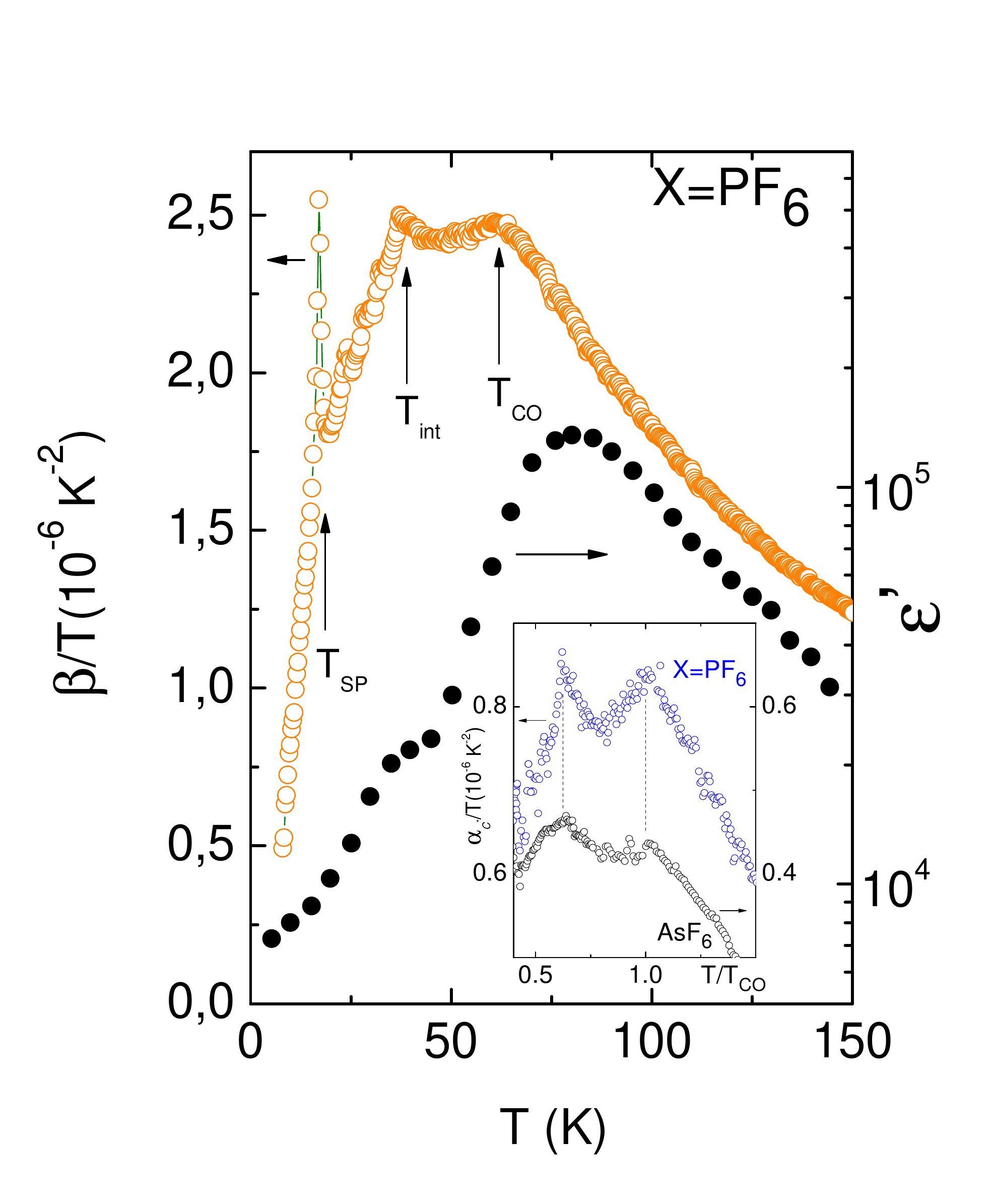}
\end{center}
\caption{\small Main panel, left scale: Volumetric thermal expansion
$\beta$ = $\alpha_a$ + $\alpha_b'$ + $\alpha_c^*$ over $T$ for
(TMTTF)$_2$PF$_6$, obtained from the data set shown in
Fig.\,\ref{TE-PF6}. Arrows indicate $T_{SP}$, $T_{int}$ and $T_{CO}$ transition temperatures. Right scale: Real part of the
dielectric constant $\epsilon'$, plotted on the same temperature
scale. Data taken from Ref.\,\cite{Nad2006-20}. Inset:
$\alpha_c^*$/$T$ versus $T/T_{CO}$ data for (TMTTF)$_2$X with X =
PF$_6$ (left scale) and X = AsF$_6$ (right scale), showing that
$T_{CO}$ and $T_{int}$ are linked by the relation $T_{int}$ $\simeq$
0.6 $\cdot$ $T_{CO}$. Adapted from
Refs.\,\cite{Mariano2008,Thesis}.}\label{BetaoverTvsepsilon}
\end{figure}

Upon cooling below $T_{CO}$, a kink is observed in the uniaxial thermal expansion coefficient at $T_{int}$ $\simeq$ (39 $\pm$ 2)\,K and (65 $\pm$ 3)\,K for X = PF$_6$ and AsF$_6$, respectively. The observed kink in $\alpha_i(T)$ at
$T_{int}$, again more pronounced along the \emph{c$^*$}-axis,
indicates the existence either of  an additional phase transition or of a crossover in the intermediate temperature range $T_{SP}
< T_{int} < T_{CO}$. These anomalous features can be better visualized in Fig.\,\ref{BetaoverTvsepsilon}, where  the volumetric thermal
expansion coefficient $\beta$, divided by $T$ for (TMTTF)$_2$PF$_6$,
is shown together with dielectric constant data extracted from the
literature \cite{Nad2006-20} for the same salt. Very sharp kinks in
$\beta(T)/T$ are observed at $T_{CO}$ and $T_{int}$. Note that the
shape of the anomalies is very similar, indicating therefore that
both features are likely to have the same origin.
Interestingly enough, a small bump at $T_{int}$ can also be observed
in the real part of the dielectric constant $\epsilon'(T)$, a
feature which has been overlooked so far.

The analogy between the anomalies at $T_{CO}$ and $T_{int}$ is
better illustrated by the inset of Fig.\,\ref{BetaoverTvsepsilon},
where $\alpha_{c^*}/T$ is plotted as a function of $T/T_{CO}$ for
the X = PF$_6$ and AsF$_6$ salts. A remarkable phenomenological
result obtained from this plot is that the anomalies at $T_{int}$
and $T_{CO}$ for both salts are linked by $T_{int} \simeq 0.6 \cdot
T_{CO}$. Interestingly enough, dielectric measurements as a function
of $T$ on the mixed-stack charge-transfer salt TTF-CA, which is
recognized as a  prototype system exhibiting the NI transition,
reveal a similar behavior \cite{Okamoto1991-20}. Upon cooling TTF-CA through the NI transition, two distinct peaks are observed in
the real part of the dielectric constant. The authors assigned these
features to the dynamics of the NI domain-pairs and ionic domains in
the neutral lattice. Amazingly, the thermal position of the peaks is
also scaled by the factor 0.6, i.e.\ the same factor that links $T_{int}$ and
$T_{CO}$ in the (TMTTF)$_2$X salts with X = PF$_6$, AsF$_6$. This similarity indicates that
there is a common origin between these behaviors which probably relies on a somewhat similar texture of the ferroelectric medium.

It has been recently observed an enhancement of the \emph{c$^*$}-axis compressibility modulus below 40\,K (i.e. below $T_{int}$) in (TMTTF)$_2$PF$_6$ \cite{Poirier}. Such a behavior recalls the enhancement of the \emph{a}-axis Young modulus observed below 55\,K in (TMTSF)$_2$PF$_6$  \cite{Chaikin1982} and recently attributed  to the H bond linkage of the PF$_6$ to the methyl groups \cite{Foury2013}. It is thus tempting to suggest that a similar linkage of the PF$_6$ to the TMTTF should occur at $T_{int}$ in (TMTTF)$_2$PF$_6$. In this scenario, structural modifications brought by this linkage should modify the CO pattern on the TMTTF. In this respect it has been observed a change of the $^{13}$C NMR spectra below about 40\,K in (TMTTF)$_2$PF$_6 ($\cite{Naka2007}). However the optical spectra do not exhibit sizeable modifications around $T_{int}$\cite{Dressel2012}.
The thermal dependence of the expansion coefficient of the SbF$_6$ salt measured along the \emph{c$^*$} direction (Fig.\,14)  however does not reveal any clear additional anomaly between its broad maximum located around 50-90\,K and $T_{CO}$. But the dielectric constant of this salt exhibits a small bump at about 100\,K (Fig.\,14) which recalls the one observed at $T_{int}$ in the PF$_6$ salt (Fig.\,16).
In conclusion, much work is necessary to determinate if there is any relation, via a modification of the CO pattern, between the various anomalies revealed by dielectric measurements and the static and dynamic properties of the anion sublattice essentially probed by thermal expansion measurements.

\section{Anomalous lattice properties of the Fabre salts in relation with the anion sublattice}
The thermal dependence of the uniaxial expansivity measurements
performed in the Fabre salts with octahedral anions exhibits a
negative slope at high temperatures. This negative contribution is
observed whatever the direction of measurement.  However this effect
is the most pronounced for measurements performed along the
interlayer direction \emph{c*} where  $\alpha_{c*}$($T$) begins to
decrease upon heating from about 75\, (PF$_6$ and SbF$_6$) -- 100\,K
(AsF$_6$). Hence, this negative contribution may indicate some kind
of unconventional lattice dynamics in the Fabre salts, which can
be understood within the ``rigid-unit mode'' (RUM) scenario,
introduced by Goodwin \emph{et al.} \cite{Goodwin2005}. In this
scenario, the thermal population of local (dispersion-less)
low-energy vibrational modes of independent rigid units which
are coordinated in a flexible fashion induces a shrinking of their
surrounding (especially in the direction perpendicular to the RUM
displacement). In this framework it is tempting to suggest that
rotational or translational modes of rigid PF$_6$, AsF$_6$ or
SbF$_6$ units trapped in centrosymmetrical anion cavities delimited
by the methyl groups \cite{Pouget96, Crystals}, could play the role
of these RUM and cause a negative contribution to $\alpha_i$. In
this respect it has been recently determined that low temperature
rotation of PF$_6$ in (TMTSF)$_2$PF$_6$ can be described by a
collection of low frequency ($\sim$45\,cm$^{-1}$) Einstein
oscillators \cite{Foury2013}. Furthermore, $^{19}$F NMR studies
\cite{Mc1982,Furukawa-2005,Yu2004-20} were able to show that
octahedral anions are disordered at high temperatures by thermal
activated jump over potential height between RT and $\sim$135\,K and
by rotation above $\sim$70\,K. In particular, measurements performed
in (TMTTF)$_2$SbF$_6$ \cite{Furukawa-2005},\cite{Yu2004-20} show
that anions rotate in the temperature range where the $c^*$ thermal
expansion coefficient exhibits a negative contribution (see
Fig.\,14). In addition, Fig.\,14 shows that $\alpha_{c^*}(T)$ is
drastically reduced below 50\,K when the anion is frozen in its
cavity.

Our interpretation is also corroborated by the fact that the magnitude of the thermal expansion coefficient is drastically reduced and its negative contribution is suppressed by substitution of octahedral anions (X = PF$_6$, AsF$_6$, SbF$_6$) by the Br \cite{Dani}  which does not possess rotational degrees of freedom.
This observation is complemented by a recent study \cite{Foury2013} of the thermal dependence of the lattice expansivity of (TMTSF)$_2$PF$_6$  below 200\,K, which  exhibits also high temperature negative contributions \cite{Lang2004},  and whose analysis shows that anion rotation exhibits a huge Gr\"uneisen parameter  of $\Gamma$(PF$_6$) $\approx$ 28. A similar value of $\Gamma$(PF$_6$) is expected in (TMTTF)$_2$PF$_6$ because the thermal expansion coefficients have the same magnitude in (TMTSF)$_2$PF$_6$ and (TMTTF)$_2$PF$_6$.\linebreak

In general, such RUM are likely the driving force of the negative thermal expansion
(NTE) phenomenon observed in several materials \cite{Goodwin2008-20g1a,
Goodwin2005,Thesis}.
As pointed out above, the CO transition seems to
affect dramatically the overall behavior of the $\alpha_{c^*}(T)$
expansivity. For the X = PF$_6$ and AsF$_6$ salts,  $T_{CO}$
coincides roughly with the temperature below which the negative
contribution to $\alpha_{c^*}(T)$ is no longer active. Above
$T_{CO}$, CO pre-transitional fluctuations cause positional
fluctuations of the anions towards their new off-center equilibrium
position, providing an effective damping/softening of these modes.
Upon achieving the structural distortion at $T_{CO}$ the anion motions
become strongly reduced which suppress the negative
contribution in $\alpha_{c^*}$ for $T$ $<$ $T_{CO}$.
In fact, data obtained for the
SbF$_6$ salt shows that the situation could be a little bit more
subtle than the one previously described. Figure 14 shows indeed that the
thermal expansion coefficient along $c^*$ has a negative slope above
$T_{CO}$, but also that the negative dependence continues below
$T_{CO}$ until about 85\,K. In the RUM scenario, this means that
the anion fluctuations do not really stop at the CO transition of
the SbF$_6$ salt, a statement in perfect agreement with $^{19}$F NMR
studies performed in (TMTTF)$_2$SbF$_6$
\cite{Yu2004-20,Furukawa-2005}. Thus, a more elaborated scenario should be
that the Fabre-Bechgaard salts possess an incipient off-centering
anion instability and that the 4\,$k_F$
CDW or CO electronic instability uses this structural instability to
stabilize at $T_{CO}$ a 3D pattern of localized charges on one site
out of two (Fig.\,1 (a) and Fig.\,17).  The non-freezing of all the anion degrees of freedom at $T_{CO}$  is sustained by the experimental observation that in the Fabre salts incorporating tetrahedral anions, such as ReO$_4$ and BF$_4$ (whose orientation is  disordered at RT), the CO transition leaves unaffected the orientation degree of freedom of the anion which thus remains available for their ordering at an AO transition with a $T_{AO}$  \cite{Pouget96} significantly  lower than $T_{CO}$ \cite{Nad2006-20}.\linebreak

In the CO process, however, the size of the
anion should also play an important
role. As discussed in
Ref.\,\cite{Pouget2007}, for anions of same symmetry
$T_{CO}$ increases with the anion size. For example,
$T_{CO}^{X = SbF_6}$ = 154\,K $>$ $T_{CO}^{X = AsF_6}$ = 105\,K
$>$ $T_{CO}^{X = PF_6}$ = 65\,K. This can be simply rationalized if the anion size defines
its free rotational/translational volume in the methyl group cavity as well as the intensity of the
electron-anion coupling    necessary to stabilize the CO pattern.\linebreak


 The off-centering anion displacements accompanying the CO transition
has been presented in Ref.\,\cite{Brazo2008}, as a direct consequence of the Earnshaw's theorem,
which states that a classical system of point charges under the
interaction of Coulomb forces alone is unstable, since there is no
minimum in the electrostatic potential. This statement however requires many caution because if the shift of the anion induces some chemical bonding, with the methyl groups for example (see the end of this section) we are faced with a quantum phenomenon going beyond the validity of the  Earnshaw's  theorem.

\begin{figure}
\begin{center}
\includegraphics[angle=0,width=0.44\textwidth]{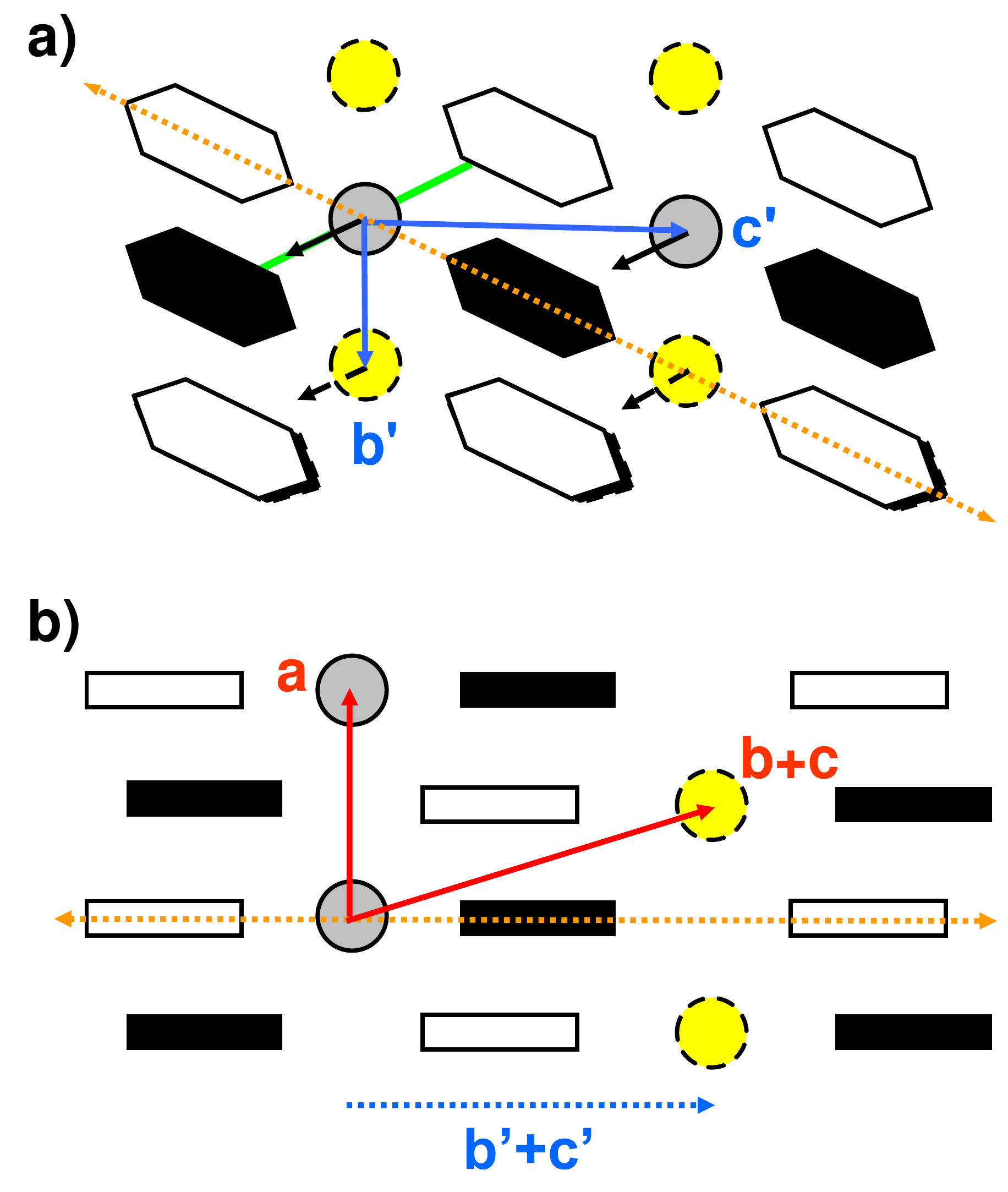}
\end{center}
\caption{\small Schematic representation of the 3D ferroelectric
pattern in the (TMTTF)$_2$X family. a) Charge pattern in
(\emph{b'}, \emph{c'}) plane. Black and white hexagons are used to
represent charge rich- and charge poor-sites, i.e.\ (TMTTF)$^{0.5 +
\rho}$ and (TMTTF)$^{0.5 - \rho}$, respectively. Dashed black
hexagons are used to represent (TMTTF)$^{0.5 + \rho}$ molecules
located in another layer. Anions are represented by full and dashed
circles. Arrows indicate displacements of the anions from their
centro-symmetrical positions towards positive charged
nearest-neighbor (TMTTF)$^{0.5 + \rho}$ molecules, necessary for the
stabilization of the CO phase. Green line exemplarily indicates
short $S$-$F$ contacts. b) View of the ferroelectric pattern in the (a,
\emph{b'} + \emph{c'}) plane, indicated by orange dotted line in a).
Adapted from Refs.\,\cite{Mariano2008,Thesis}.}\label{FE-3D}
\end{figure}

The present findings (see Figs.\,\ref{TE-PF6}, \ref{TE-AsF6} and
\ref{SbF6}) provide strong evidence that the interstack
\emph{$c^*$}-axis plays a crucial role for the stabilization of
the CO phase. The \emph{c$^*$} direction is distinct in that it
incorporates the anions X (see Figs.\,\ref{TMTTFstr} and
\ref{TMTTFstr_bc}), while the \emph{a}-axis lattice parameter,
which is determined by the intrastack interactions between
neighboring TMTTF molecules, remains practically unaffected by the CO, cf.\
Fig.\,\ref{TE-PF6}. Hence, according to the anisotropy observed in
$\alpha_i$ with dominant effect in $\alpha_{c^*}$, anions
displacement and their coupling to the TMTTF molecules play a
crucial role in the CO transition, as previously predicted
\cite{Riera2001}. Such a scenario, taking into account both the charge degrees of
freedom on the TMTTF stacks as well as
their coupling to the anions,  is illustrated in Fig.\,\ref{FE-3D}
\cite{Mariano2008}.
In this figure the arrangement of TMTTF (hexagons) and the anions
(circles) in the (\emph{b'}, \emph{c'}) plane (top panel) and the
(\emph{a}, \emph{b'} + \emph{c'}) planes (bottom panel) are shown
schematically. Upon cooling through $T_{CO}$, the charge $\rho$ on
the TMTTF molecule changes from a homogeneous distribution with
$\rho$ = 0.5 (in units of $e$) hole per TMTTF, above $T_{CO}$ to a
modulated structure whose charges alternate by $\pm \delta$ along
the TMTTF stacks below. For deriving the resulting 3D charge
pattern, we start by considering a stack of anions along the
$\textbf{\textit{a}}$-axis and the two nearest-neighbor stacks of
TMTTF molecules linked via short S-F contacts (green line in
Fig.\,\ref{FE-3D}). For a fixed charge modulation on one of the
stacks, the electrostatic energy of the whole array can be reduced
if one of the anions' nearest-neighbor TMTTF molecules is
charge-rich (TMTTF)$^{(\rho_{0} + \delta)}$ (black symbols), the
other one charge poor (TMTTF)$^{(\rho_{0} - \delta)}$ (white
symbols), while the anions perform slight shifts towards the
charge-rich molecule. The resulting anion displacements, indicated
by the arrows in Fig.\,\ref{FE-3D}, which are uniform for all anions
and lift the inversion symmetry, together with the minimization of
Coulomb energies of adjacent stacks along the
$\textbf{\textit{b}}$-axis determine the 3D charge pattern
unambiguously. \newline Thus, the CO transition achieving
ferroelectricity requires a shift of the anions from the high
temperature inversion centres. The direction of shift of the anions
has not yet been determined for the Fabre salts,  but is known in
the parent salt  $\delta$-(EDT-TTF-CONMe$_2$)$_2$Br
\cite{Zorina2009}. In this respect, two possible scenarios can be
considered \cite{Crystals,Pouget-PPS2012}:  either\linebreak
\textbf{(1)} a shift of the anion towards the S atom of the TMTTF,
shortening one S-F contact distance (this is the scenario
illustrated in Fig.\,17) or\linebreak \textbf{(2)} a shift of the
anion towards the methyl groups of one of the TMTTF molecules
delimiting the methyl group cavity, forming F...H-CH$_2$ bonds with
this molecule.\linebreak In each case the anion shift stabilizes an
excess of hole in the molecule towards which the anion moves.
However, the location of charge-rich TMTTF's is different in these
two scenarios because the direction of displacement of the anion is
different. Arguments in favor of scenario (2) are given in
Ref.\,\cite{Crystals}. In this last scenario, the shift of the anion
and its H-bonding with the methyl groups of one TMTTF out of two
will form a H-bond ferroelectric \cite{Kittel}. H-bond
ferroelectricity explains both the strong H/D isotopic effect
leading to a significant increase of $T_{CO}$ when D substitutes H
in TMTTF \cite{Pouget2007} and the order-disorder,  or
relaxational-type, dynamics of the ferroelectric transition recently
revealed by cleaver dielectric measurements \cite{Staresinic-2006}.
In the order-disorder scenario, the two relevant configurations
correspond to the two manners that the MF$_6$ anion has to form
F...H bond with the neighboring methyl groups. As methyl groups
rotate easily, this ferroelectricity should be very sensitive to
defects. Disorder of the methyl groups will perturb the H bonding
network with the anions. Thus an opposite F...H-CH$_2$ linkage  will
invert the direction of the ferroelectric polarization and, since
F...H-bonds are directed towards the hole rich TMTTF, this  will
induce a phase shift in the CO pattern. Disorder will thus fragment
the ferroelectric pattern into domains which achieve local
polarization at the origin of the relaxor-like dielectric
properties. We suspect that the PF$_6$ salt having the lowest
$T_{CO}$ should be more sensitive to the disorder because $T_{CO}$
is close to the freezing point of the rotational classical motion of
the methyl groups ($\sim$55\,K in (TMTSF)$_2$PF$_6$
\cite{McBrierty1982,Scott1982}). In addition, X-ray irradiation
defects will easily disorder the H-bond pattern leading to a
disordered CO ground state. The break of the long-range CO explains
why X-ray diffraction experiments have not succeeded to reveal the
symmetry breaking at $T_{CO}$, and why neutron scattering, which
does not induce methyl group disorder, has recently succeeded to
detect a singularity in the thermal dependence of the Bragg
reflection intensity at $T_{CO}$  \cite{Foury2010,Pouget-PPS2012}.

\section{Summary and Outlook}
In this review we have discussed several aspects of the CO transition.

The thermal expansion measurements reviewed here revealed, for the first time, lattice effects associated with
the  CO transition  of (TMTTF)$_2$X salts. Given the
charge unbalancing imposed by the CO transition,
uniform displacements of negative charged anions X$^-$ from their
symmetric positions towards positive charged nearest-neighbor
(TMTTF)$^{0.5 + \rho}$ are required to minimize the electrostatic
energy. Hence, expansivity results demonstrate the importance of
lattice effects/anion displacements for the stabilization of the
CO phase, as predicted from calculations
on the basis of the extended Hubbard model including Peierls-like
coupling. The anomalous expansivity observed above about 100\,K along the \emph{c$^*$}-axis,
along which planes of TMTTF molecules are separated  by planes of
anions X was interpreted
considering the rigid-unit-mode scenario. In (TMTTF)$_2$X salts (X = PF$_6$ and
AsF$_6$), the thermal expansion coefficient along the
\emph{c$^*$}-axis remains positive while a dramatic change of slope
occurs from $d \alpha_{c^*} / dT$ $<$ 0 above $T_{CO}$ to $d \alpha_{c^*}
/ dT$ $>$ 0 below $T_{CO}$.
Although the CO transition in (TMTTF)$_2$X exhibits  a mean-field behavior (the dielectric constant
obeys the Curie law), charge-ordering fluctuations show up well
above $T_{CO}$. Such
charge-fluctuations couple to the rotational or
translational shifts of the rigid PF$_6$ or AsF$_6$ units from the
center of the cavities delimited by the methyl groups.
Below the CO transition temperature ($T_{CO}$ = 65\,K and
105\,K for X = PF$_6$ and AsF$_6$ salts, respectively),
translations or rotations of the anions are no longer
critically active and, as consequence, a dramatic change of slope in the
thermal expansion coefficient along the \emph{c$^*$}-direction is
observed in these salts. Based on the strong anisotropy observed, a scheme to
demonstrate the 3D ferroelectric character of the
CO transition, taking into account charge-degrees of freedom and
anion displacements, was proposed. Furthermore, evidence for a new
phase transition or crossover were presented for the PF$_6$ and AsF$_6$ salts at $T_{int}$
$\simeq$ 0.6 $\cdot$ $T_{CO}$. This unexpected finding could be caused by the linkage of the anion to the methyl groups in the same temperature range for (TMTSF)$_2$PF$_6$.
Measurements along the
\emph{c$^*$}-axis on the (TMTTF)$_2$X salt with X = SbF$_6$
revealed a distinct behavior. The shape and magnitude of the lattice expansion
anomaly, indicative of the presence of strong critical fluctuations, differ
markedly from the mean-field-like anomaly observed at $T_{CO}$ for the X =
PF$_6$, AsF$_6$ salts. This difference can be understood as a
consequence of short-range Coulomb forces in the X = SbF$_6$ salt \cite{Dani}, where
CO coincides with a metal-to-insulator transition, i.e. $T_{CO}$ =
$T_{MI}$. This  implies a more effective screening of the long-range
Coulomb forces above $T_{CO}$, as compared to the X = AsF$_6$ and PF$_6$
salts, where $T_{CO} < T_{\rho}$, with $T_{\rho}$ denoting the temperature of minimum of resistivity, which marks the onset of
the charge localization \cite{Laversanne84,Nad2006-20,Coulon82}.\linebreak

In what follows, we suggest further experiments which will contribute to shed light into the CO phase.
For the (TMTTF)$_2$X salts (TMTTF)$_2$X salts high-resolution thermal expansion measurements both at the CO and AO transitions of the X = ReO$_4$ and BF$_4$ salts are desirable. In addition,  directional-dependent thermal
expansion measurements on (TMTTF)$_2$SbF$_6$ are important in
order to look for anisotropic effects at the CO
transition which in turn is accompanied by a Mott
metal-to-insulator transition. Systematic measurements on the TMTSF family are important in order to check the existence of  a possible ``metallic ferroelectricity'' as suggested long time ago by Anderson and Blount \cite{Anderson} or of an
incipient local and fluctuating ferroelectricity \cite{Brazo2008}. We have shown in this review that the anions play an important role in stabilizing the CO pattern. In the same spirit it should be
important to precise more clearly their influence in the AO microscopic mechanism especially in compound such as (TMTSF)$_2$ClO$_4$ where their kinetic of ordering tunes the electronic ground state. In particular, the effect of elastic deformations of the conducting layers induced by the AO process should be clarified.  In this respect, it should be important to perform high-resolution lattice parameter measurements in (TMTSF)$_2$ClO$_4$
around the AO transition in order to complete earlier measurements which already show an influence of the cooling rate on the \emph{a} \cite{Goanach1987}, $\gamma$ and \emph{c} \cite{Fertey94,Fertey95}
triclinic parameters (see also Ref.\,\cite{Crystals}). Such measurements could also reveal whether there exist similarities between the AO and the glass-like
transition observed in fully deuterated/hydrogenated salts of
$\kappa$-(BEDT-TTF)${_2}$Cu[N(CN)$_{2}$]Br \cite{Thesis}, and how an inhomogeneous texture of the low temperature
phases of these organic salts  could explain the strong sensitivity of their superconducting properties to cooling effects\cite{Haddad}.

In order to gain more insights into the screening and polarization effects in the (TMT$C$F)$_2$X salts, systematic calculations of the Madelung energy \cite{Nogueira} above and
below the CO transition temperature, including the polarization effects induced by the anion shift, are required. Furthermore, a detailed experimental investigation of the recent proposed multiferroic
character of (TMTTF)$_2$PF$_6$ \cite{Gianluca}, similar to the one proposed for TTF-CA \cite{Joren}, is highly desired as well.\newline

\section*{Acknowledgements}
The thermal expansion results reported here has been the subject of a long term collaboration as can be seen in the references list.
Special acknowledgements are due to M. Lang and P. Foury-Leylekian for fruitful discussions along the years.
MdS acknowledges financial support from the S\~ao Paulo
Research Foundation -- Fapesp (Grants No. 2011/22050-4) and National Counsel of Technological and Scientific Development -- CNPq (Grants No. 308977/2011-4).


\normalsize

\begin{thebibliography}{99}

\bibitem{Imada} M. Imada, A. Fujimori and Y. Tokura,
Rev. Mod. Phys. \textbf{70}, 1039 (1998).

\bibitem{Spiral} M. Mostovoy, Phys. Rev. Lett. \textbf{96}, 067601
(2006).

\bibitem{Magneto} H. Katsura,  N. Nagaosa and A.V. Balatsky, Phys.
Rev. Lett. \textbf{95}, 057205 (2005).

\bibitem{Seo} H. Seo,  J. Merino,  H. Yoshioka and M. Ogata, J. Phys. Soc. Jpn. \textbf{75}, 051009 (2006).

\bibitem{Kakiuchi} T. Kakiuchi, Y. Wakabayashi, H. Sawa, T. Itou and K. Kanoda, Phys. Rev. Lett. \textbf{98}, 066402 (2007).


\bibitem{Pouget 1976} J.-P. Pouget, S.K.Khanna, F. Denoyer, R.Comes, A.F. Garito and A.J. Heeger, Phys. Rev. Lett. \textbf{75}, 437 (1976).

\bibitem{Hubbard1978} J. Hubbard, Phys. Rev. B \textbf{17}, 494
(1978).

\bibitem{Kondo}  J. Kondo and K. Yamaji, J. Phys. Soc. Jpn. \textbf{43}, 424 (1977).


\bibitem{Pouget2012} J.-P. Pouget, Physica B \textbf{407}, 1762 (2012).

\bibitem{Lunke} P. Lunkenheimer, J. M\"uller, S. Krohns, F. Schrettle, A. Loidl, B. Hartmann, R. Rommel, M. de Souza, C. Hotta, J.A. Schlueter and M. Lang, Nature Materials \textbf{11}, 755  (2012).


\bibitem{Lang4} N. Toyota, M. Lang and J. M\"uller, Low-Dimensional
Molecular Metals, Springer, Germany (2007).

\bibitem{Lang20a} M. Lang, Superconductivity Review
\textbf{2}, 1 (1996).


\bibitem{Ishiguro20} T. Ishiguro and K. Yamaji, \emph{Organic
Superconductors} in Springer Series in Solid-State Sciences 88,
Springer-Verlag, Germany (1990).

\bibitem{Jerome1991-20} D. J\'erome, The Physics of Organic
Superconductors, Science \textbf{252}, 1509 (1991).


\bibitem{Mckenzie1998-20} R.H. McKenzie, Comments
Condensed Matter Physics \textbf{18}, 309-337 (1998).

\bibitem{Wosnitza2000-20} J. Wosnitza, Studies of High Temperature Superconductors \textbf{34}, Nova Science Publishers,
97-131 (2000).

\bibitem{Singleton2002-20} J. Singleton, Journal of Solid State
Chemistry \textbf{168}, 675 (2002).

\bibitem{Myagawa2004-20} K. Miyagawa, K. Kanoda and A. Kawamoto,
Chem. Rev. \textbf{104}, 5635-5653 (2004).

\bibitem{Fukuyama2006-20} H. Fukuyama, J. Phys. Soc. Jpn. \textbf{75}, 051001-1 (2006).

\bibitem{Mori2006-20} H. Mori, J. Phys. Soc. Jpn. \textbf{75}, 051003-1 (2006).

\bibitem{Kanoda2006-20} K. Kanoda, J. Phys. Soc. Jpn. \textbf{75}, 051007-1 (2006).

\bibitem{Powell2006-20} B.J. Powell and R.H. McKenzie, J. Phys.:
Condens. Matter \textbf{18}, R827-R866 (2006).

\bibitem{Dressel2007g7} M. Dressel, Naturwissenschaften \textbf{94}, 527 (2007).

\bibitem{Lebed} The Physics of Organic Superconductors and Conductors,
edited by A. Lebed, Springer Series in Materials Science
Vol. 110, Springer-Verlag, Berlin, 2008.

\bibitem{Monceau2001-20} P. Monceau, F.Ya. Nad and S. Brazovskii,
Phys. Rev. Lett. \textbf{86}, 4080 (2001).

\bibitem{Gianluca} G. Giovannetti, S. Kumar, J.-P. Pouget and M. Capone, Phys. Rev. B \textbf{85}, 205146 (2012).

\bibitem{Crystals} J.-P. Pouget, Crystals  \textbf{2}, 466 (2012).


\bibitem{Laversanne84} R. Laversanne, C. Coulon, B. Gallois, J.-P. Pouget and R. Moret, J. Phys. Lett.
(France) \textbf{45}, L393 (1984).

\bibitem{Onsite} The on-site
Coulomb repulsion can be seen as the energy that would be necessary
to place 2 electrons of opposite spins in the same molecular
orbital.

\bibitem{Mila}  F. Mila and X. Zotos, Europhysics Letters \textbf{24}, 133 (1993).

\bibitem{Clay}  R.T. Clay, S. Mazumdar and D.K. Campbell, Phys. Rev. B \textbf{67}, 115121 (2003).

\bibitem{Emery88} V.J. Emery and C. Noguera, Phys. Rev. Lett.  \textbf{60}, 631 (1988).

\bibitem{BOW} For an uniform chain, the bonds linking first neighbour sites in the 4$k_F$ BOW are
fluctuating in a resonant manner. Resonant fluctuations are suppressed if an
incipient lattice dimerization pins the bonds of the 4$k_F$ BOW on dimers. The static
order which thus results is generally referred to as a MD ground state in the
literature. This is the pinned state which is represented in Fig.\,1 (c).



\bibitem{Tsuchiizu2001}  M. Tsuchiizu, H. Yoshioka and Y. Suzumura, J. Phys. Soc. Jpn \textbf{70}, 1460 (2001).

\bibitem{Riera2001} J. Riera and D. Poiblanc, Phys. Rev. B
\textbf{63}, 241102-1 (2001).

\bibitem{Moldenhauer9a1} J. Moldenhauer, Ch. Horn, K. I. Pokhodnia and D. Schweitzer, Synthetic
Metals \textbf{60}, 31 (1993).

\bibitem{Hiraki9a2} K. Hiraki and K. Kanoda, Phys. Rev. Lett. \textbf{80}, 4737 (1998).

\bibitem{Ilakovac} V. Ilakovac, S. Ravy, A. Moradpour, L. Firlej and P. Bernier, Phys. Rev. B \textbf{52}, 4108 (1995).

\bibitem{Gallois20g5} B. Gallois, Ph.D.\ Thesis (in French), University of Bordeaux
(1987).

\bibitem{Chasseaug20g6} D. Chasseau, J. Gaultier, J. L. Miane, C. Coulon, P.
Delhaes, S. Flandrois, J. M. Fabre and L. Giral, J. Phys. Colloq.
France \textbf{44}, C3-1223 (1983).

\bibitem{Furukawa-2005} K. Furukawa, T. Hara and T. Nakamura, J. Phys. Soc. Jpn. \textbf{74}, 3288 (2005).

\bibitem{Kistenmacher1984} T.J. Kistenmacher , Solid State Commun. \textbf{51}, 931 (1984).

\bibitem{Scott1982} J.C. Scott, E.M. Engler, W.G. Clark, C. Murayama, K. Bechgaard and H.J. Pedersen, Mol. Cryst. Liq. Cryst. 79, 61 (1982).

\bibitem{McBrierty1982} V.J. McBrierty, D.C. Douglass, F. Wudl and E.
Aharon-Shalom, Phys. Rev. B \textbf{26}, 4805 (1982).

\bibitem{Mc1982} V.J. McBrierty, D.C. Douglass and F. Wudl, Solid State Commun. \textbf{43}, 679 (1982).

\bibitem{Yu2004-20} W. Yu, F. Zhang, F. Zamborszky, B. Alavi, A.
Baur, C.A. Merlic and S.E. Brown, Phys. Rev. B \textbf{70}, 121101
(2004).

\bibitem{Gallois86} B. Gallois, J. Gaultier, C. Hauw, T. Lamcharfi and A. Filhol, Acta Cryst. B \textbf{42}, 564 (1986).

\bibitem{Foury2013} P. Foury-Leylekian, S. Petit, I. Mirebeau, G. Andre, M. de Souza,  M. Lang, E. Ressouche, A. Moradpour and J.-P. Pouget, Phys. Rev. B \textbf{88}, 024105 (2013).

\bibitem{Granier1988} T. Granier, B. Gallois, L. Ducasse, A. Fritsch and A. Filhol, Synthetic Metals \textbf{24}, 343 (1988).

\bibitem{Pouget96} J.-P. Pouget and S. Ravy, J. Phys. I (France)
\textbf{6}, 1501 (1996).

\bibitem{Emery82} V.J. Emery, R. Bruinsma and S. Barisic, Phys. Rev. Lett. \textbf{48},  1039 (1982).

\bibitem{Coulon82} C. Coulon, P. Delhaes, S. Flandrois, R. Lagnier, E. Bonjour and J.M. Fabre, J. Physique \textbf{43}, 1059 (1982).

\bibitem{Jerome82} D. J\'erome, A. Mazaud, M. Ribault and K.
Bechgaard, J. Phys. Lett. \textbf{41}, L95 (1980).


\bibitem{Bourbonnais 97} C. Bourbonnais, Synthetic Metals \textbf{84}, 19 (1997).

\bibitem{Bourbonnais 98}  C. Bourbonnais and D. J\'erome, Science \textbf{281}, 1155 (1998).

\bibitem{Adachi2000} T. Adachi, E. Ojima, K. Kato
and H. Kobayashi, J. Am. Chem. Soc. \textbf{122}, 3238 (2000).

\bibitem{foot1} A similar model to treat the orientational degrees of freedom of
the ethylene end-groups of the ET molecules in the quasi-2D organic
(ET)$_2$X charge-transfer salt family was proposed by S. Ravy
\emph{et al.} (S. Ravy, R. Moret and J.-P. Pouget, Phys. Rev. B
\textbf{38}, 4469 (1988)) for the $\beta$-phase and by J.\ M\"uller
(J. M\"uller, Ph.D. Thesis (in German), Max-Planck-Institute,
Dresden, Germany (2001)) for the $\kappa$-phase. See also J.-P.
Pouget Mol. Cryst. Liq. Cryst. \textbf{230}, 101 (1993).

\bibitem{Coulon82a} C. Coulon, A. Maaroufi, J. Amiell, E. Dupart, S. Flandrois, P. Delhaes, R. Moret, J.-P. Pouget and J.P. Morand, Phys. Rev. B \textbf{26}, 6322 (1982).

\bibitem{Emery83}  V.J. Emery, J. Phys. Coll. \textbf{44} C3, 977 (1983).

\bibitem{Brazovskii85} S. Brazovskii and V. Yakovenko, J. Phys. Lett. (Paris) \textbf{46}, L111 (1985).

\bibitem{Coulon85}  C. Coulon, S.S.P. Parkin, R. Laversanne, Mol. Crys. Liq. Cryst. \textbf{119}, 325 (1985).

\bibitem{Javadi88} H.H.S. Javadi, R. Laversanne and A.J. Epstein,
Phys. Rev. B \textbf{37}, 4280 (1988).

\bibitem{Chow9a3} D.S. Chow , F. Zamborszky, B. Alavi, D.J. Tantillo, A. Baur, C.A. Merlic,
and S. E. Brown, Phys. Rev. Lett. \textbf{85}, 1698 (2000).

\bibitem{Nad2000} F. Nad, P. Monceau, C. Carcel and J.M. Fabre,
Phys. Rev. B \textbf{62}, 1753 (2000).

\bibitem{Dumm05-20} M. Dumm, M. Abaker and M. Dressel, J. Phys. IV
France \textbf{131}, 55 (2005).

\bibitem{Menegheti1984} M. Meneghetti, R. Bozio, I.
Zanon, C. Pecile, C. Ricotta and M. Zanetti, J. Chem. Phys.
\textbf{80}, 6210 (1984).

\bibitem{Foury2010} P. Foury-Leylekian, S. Petit, G. Andre, A. Moradpour and J.-P. Pouget,
Physica B \textbf{405}, S95 (2010).

\bibitem{Pouget-PPS2012} J.-P. Pouget, P. Foury-Leylekian, P. Alemany and E. Canadell,
Phys. Status Solidi B \textbf{249}, 937 (2012).

\bibitem{Coulon-2007} C. Coulon, G. Lalet, J.-P. Pouget, P. Foury-Leylekian, A. Moradpour and J.-M. Fabre, Phys. Rev. B \textbf{76}, 085126 (2007).

\bibitem{Pouget2007} J.-P. Pouget, P. Foury-Leylekian, D. Le
Bolloc'h, B. Hennion, S. Ravy, C. Coulon, V. Cardoso and A.
Moradpour, J. Low Temp. Phys. \textbf{142}, 147 (2007).

\bibitem{Zorina2009} L. Zorina, S. Simonov, C. Meziere, E. Canadell, S. Suh, S.E. Brown, P. Foury-Leylekian, P. Fertev, J.-P. Pouget and P. Batail, J. Mater. Chem. \textbf{19}, 6980 (2009).

\bibitem{Dumm2000} M. Dumm, A. Loidl, B.W. Fravel, K.P. Starkey, L.K.
Montgomery and M. Dressel, Phys. Rev. B \textbf{61}, 511 (2000) and
earlier references therein.

\bibitem{Zamborsky2002-20} F. Zamborszky, W. Yu, W. Raas, S.E. Brown,
B. Alavi, C.A. Merlic and A. Baur, Phys. Rev. B \textbf{66}, 081103
(2002).

\bibitem{Langlois} A. Langlois, M. Poirier, C. Bourbonnais, P. Foury-Leylekian, A. Morapour and J.-P. Pouget, Phys. Rev. B \textbf{81}, 125101 (2010).

\bibitem{Iwase2011} F. Iwase, K. Sugiura, F. Furukawa and T. Nakamura, Phys. Rev. B \textbf{84}, 115140 (2011).

\bibitem{Nad2006-20} F. Nad and P. Monceau, J. Phys. Soc. Jpn. \textbf{75}, 051005 (2006).

\bibitem{Cheng1998} Z.-Y. Cheng, R.S. Katiyar, X. Yao and A.S.
Bhalla, Phys. Rev. B \textbf{57}, 8166 (1998).

\bibitem{Staresinic-2006} D. Stare\~sinic, K. Biljakovic, P. Lunkenheimer and A. Loidl, Solid State Commun. \textbf{137}, 241 (2006).








\bibitem{SL} R. S. Manna, M. de Souza, A. Bruehl, J. A. Schlueter and M. Lang, Phys. Rev. Lett. \textbf{104}, 016403 (2010).

\bibitem{Mariano2008} M. de Souza, P. Foury-Leylekian, A. Moradpour,
J.-P. Pouget and M. Lang,  Phys. Rev. Lett. \textbf{101}, 216403
(2008).

\bibitem{Mott} M. de Souza, A. Br\"uhl, Ch. Strack, B. Wolf, D. Schweitzer and M. Lang,
Phys. Rev. Lett. \textbf{99}, 037003 (2007).

\bibitem{Thesis}  M. de Souza, Ph.D. Thesis (2008) Universivity of
Frankfurt.
[http://publikationen.ub.uni-frankfurt.de/volltexte/2009/6240/].

\bibitem{Bartosch}  L. Bartosch, M. de Souza and M. Lang, Phys. Rev. Lett. \textbf{104}, 245701 (2010).

\bibitem{RSI} R. Manna, B. Wolf, M. de Souza and M. Lang, Rev. of Scientific Instr. \textbf{83}, 085111 (2012).





\bibitem{Barron9f} T.H.K. Barron , J.G. Collins  and G.K. White,
Adv. Phys.
\textbf{29}, 609-730 (1980).

\bibitem{Grueneisen1908} E. Gr\"uneisen, Annalen der Physik
\textbf{331}, 393 (1908).

\bibitem{NTE} J.S.O. Evans, J. Chem. Soc., Dalton Trans. \textbf{19}, 3317
(1999); G.D. Barrera, J.A.O Bruno, T.H.K. Barron and N.L. Allan,
J. Phys.: Condens. Matter \textbf{17}, R217 (2005).


\bibitem{Jens2000-9c} J. M\"uller, M. Lang, F. Steglich, J.A.
Schlueter, A.M. Kini, U. Geiser, J. Mohtashman, R.W. Winter, G.L.
Gard, T. Sasaki and N. Toyota, Phys. Rev. B \textbf{61}, 11739
(2000).

\bibitem{Gladstone1969-9c} G. Gladstone, M.A. Jensen and J.R.
Schrieffer: In: Superconductivity Vol. \textbf{2} ed. by R.D. Parks
(M. Dekker, New York 1969) p. 655.

\bibitem{Meingast1991-9c} C. Meingast, B. Blank, H. B\"urkle, B.
Obst, T. Wolf, H. W\"uhl, V. Selvamanickam and K. Salama, Phys. Rev.
B \textbf{41}, 11299 (1990).

\bibitem{Lang2004-9c} M. Lang and J. M\"uller, \emph{Organic
Superconductors} in \emph{The Physics of Superconductors} Vol. II,
K.H. Bennemann, J.B. Ketterson (eds.) p. 435-554 (Springer-Verlag),
Berlin (2004).

\bibitem{Dani} M. de Souza, D. Hofmann, P. Foury-Leylekian, A. Moradpour, J.-P. Pouget, M. Lang
Physica B: Condensed Matter \textbf{405}, S92 (2010).

\bibitem{foot3} Data along the \emph{a}-axis for the X = AsF$_6$ salt are missing
because the sample simply cleaved as soon as a small force was
exerted by dilatometer.

\bibitem{Imai1977} K. Imai, J. Phys. Soc. Jpn. \textbf{43},
1320 (1977).

\bibitem{Sawada1948} S. Sawada and G. Shirane, J. Phys. Soc. Jpn. \textbf{4},
52 (1949).

\bibitem{eu2009} M. de Souza, A. Bruhl, J. Muller, P. Foury-Leylekian, A. Moradpour, J.-P. Pouget and M. Lang, Physica B \textbf{404}, 494 (2009).

\bibitem{Pouget82} J.-P. Pouget,   R. Moret,  R. Comes, K. Bechgaard,  J.M. Fabre and L.
Giral, Mol. Cryst. and Liq. Cryst. \textbf{79}, 485 (1982).

\bibitem{Zamborszky-02} F. Zamborszky, W. Yu, W. Raas, S.E. Brown, B. Alavi, C.A. Merlic, A. Baur, S. Lefebvre and P. Wzieteck, J. Phys. (Paris) IV \textbf{12}, Pr9-139 (2002).

\bibitem{Lang2004} M. Lang, J. M\"uller, F. Steglich, B. Wolf, M.
Dumm and M. Dressel, J. Phys. IV (France) \textbf{114}, 111 (2004).

\bibitem{Okamoto1991-20} H. Okamoto, T. Mitani, Y. Tokura, S.
Koshihara, T. Komatsu, Y. Iwasa, T. Koda and G. Saito, Phys. Rev. B
\textbf{43}, 8224 (1991).

\bibitem{Poirier} M. Poirier, A. Langlois, C. Bourbonnais, P. Foury-Leylekian, A. Morapour and J.-P. Pouget, Phys. Rev. B \textbf{86}, 085111 (2012).

\bibitem{Chaikin1982} P.M. Chaikin, T. Tiedje and A.N. Bloch, Solid State Commun. \textbf{41}, 739 (1982).

\bibitem{Naka2007} T. Nakamura, K. Furukawa and T. Hara, J. Phys. Soc. Jpn. \textbf{76}, 064715 (2007).


\bibitem{Dressel2012} M. Dressel, M. Dumm, T.  Knoblauch and M. Masino, Crystals \textbf{2}, 528 (2012).

\bibitem{Goodwin2005} A.L. Goodwin and C.J. Kepert, Phys.
Rev. B \textbf{71}, 140301 (2005).

\bibitem{Goodwin2008-20g1a} A.L. Goodwin, M. Calleja, M.J. Conterio, M.T. Dove, J.S.O. Evans, D.A. Keen, L. Peters and M.G.
Tucker, Science \textbf{319}, 794 (2008).

\bibitem{Brazo2008} S. Brazovskii, \emph{The Physics of Organic Superconductors and Conductor}, Ed. A.G. Lebed, Springer Series in Materials Science, Vol. \textbf{110}, 313-356 (2008).

\bibitem{Kittel} Charles Kittel, Introduction to Solid State Physics, 7th Edition, p. 68, John Wiley and Sons, USA (1996).







\bibitem{Anderson} P.W. Anderson and E.I. Blount, Phys. Rev. Lett. \textbf{14}, 217 (1965).

\bibitem{Goanach1987} C. Goanach, G. Creuzet and C. Noguera, J. Phys. (Paris) \textbf{48}, 107 (1987).

\bibitem{Fertey94} P. Fertey, F. Sayetat, J. Muller, J.-P. Pouget, C. Lenoir and P. Batail
Physica C \textbf{235}, 2459 (1994).

\bibitem{Fertey95} P. Fertey, E. Canadell, J.-P. Pouget, F. Sayetat, C. Lenoir, P. Batail and J. Muller, Synthetic Metals \textbf{70}, 761 (1995).

\bibitem{Haddad} S. Haddad, S. Charfi-Kaddour and J.-P. Pouget
            J. Phys.: Condens. Matter  \textbf{23}, 464205 (2011).

\bibitem{Nogueira} C. Noguera, J. Phys. C: Solid State Phys.  \textbf{18}, 1647 (1985).

\bibitem{Joren} G. Giovannetti, S. Kumar, A. Stroppa, J. van den Brink  and S. Picozzi, Phys. Rev. Lett. \textbf{103}, 266401 (2009).

\end{thebibliography}
\end{document}